\documentclass[10pt,journal]{IEEEtran}
\makeatletter
\def\ps@headings{%
\def\@oddhead{\mbox{}\scriptsize\rightmark \hfil \thepage}%
\def\@evenhead{\scriptsize\thepage \hfil \leftmark\mbox{}}%
\def\@oddfoot{}%
\def\@evenfoot{}}
\makeatother
\usepackage{amssymb}
\usepackage[dvips]{graphicx}
\usepackage{subfigure}
\usepackage[lined,boxed,commentsnumbered, ruled]{algorithm2e}  
\usepackage{algorithmic}
\usepackage{amsthm}
\usepackage{mathptm}
\usepackage{cite}
\usepackage{color}
\usepackage{amsmath}

\usepackage{multicol}

\renewcommand{\theenumi}{\roman{enumi}}
\newtheorem{theorem}{Theorem}
\newtheorem{lemma}{Lemma}
\newtheorem{definition}{Definition}
\newtheorem{corollary}{Corollary}
\newtheorem{proposition}{Proposition}

\begin{document}

\title{Multi-message Authentication over Noisy Channel with Secure Channel Codes}

\author{Dajiang~Chen,~\IEEEmembership{Member,~IEEE},
~Ning~Zhang,~\IEEEmembership{Member,~IEEE},
~Nan~Cheng,~\IEEEmembership{Member,~IEEE},
 ~Kuan~Zhang,~\IEEEmembership{Member,~IEEE},
~Kan~Yang,~\IEEEmembership{Member,~IEEE},
~Zhiguang~Qin,~\IEEEmembership{Member,~IEEE},\\
and~ Xuemin (Sherman) Shen,~\IEEEmembership{Fellow,~IEEE}
\IEEEcompsocitemizethanks{
%
\IEEEcompsocthanksitem Dajiang Chen and Zhiguang Qin are with the School of information and software engineering, University of Electronic Science and Technology of China, Chengdu, 611731, China. (Email: dajiang.chen@uwaterloo.ca; qinzg@uestc.edu.cn)
\IEEEcompsocthanksitem Dajiang Chen, Ning Zhang, Nan Cheng, Kuan Zhang and Xuemin (Sherman) Shen are with the Department
of Electrical and Computer Engineering, University of Waterloo, Waterloo, ON
N2L 3G1, Canada (e-mail: dajiang.chen@uwaterloo.ca; n35zhang@uwaterloo.ca; n5cheng@uwaterloo.ca; k52zhang@uwaterloo.ca; sshen@uwaterloo.ca)
\IEEEcompsocthanksitem Kan Yang is with  the Department of Computer Science, University of Memphis, Memphis, TN 38152-6400, USA (e-mail: kan.yang@uwaterloo.ca)}
}

\maketitle

\begin{abstract}
In this paper, we investigate multi-message authentication to combat adversaries with infinite computational capacity.
An authentication framework over a wiretap channel $(W_1,W_2)$ is proposed to achieve information-theoretic security with the same key. The proposed framework bridges the two research areas in physical (PHY) layer security: secure transmission and message authentication. Specifically, the sender Alice first transmits message $M$ to the receiver Bob over $(W_1,W_2)$ with an error correction code; then Alice employs a hash function (i.e., $\varepsilon$-AWU$_2$ hash functions) to generate a message tag $S$ of message $M$ using key $K$, and encodes $S$ to a codeword $X^n$ by leveraging an existing strongly secure channel coding with exponentially small (in code length $n$) average probability of error; finally, Alice sends $X^n$ over $(W_1,W_2)$ to Bob who authenticates the received messages. We develop a theorem regarding the requirements/conditions for the authentication framework to be information-theoretic secure for authenticating a polynomial number of messages in terms of $n$.
Based on this theorem, we propose an authentication protocol that can guarantee the security requirements,
and prove its  authentication rate can approach infinity when $n$ goes to infinity.
Furthermore, we design and implement an efficient and feasible authentication protocol over binary symmetric wiretap channel (BSWC)  by using \emph{Linear Feedback Shifting Register} based (LFSR-based) hash functions and strong secure polar code.
Through extensive experiments, it is demonstrated that the proposed protocol can achieve low time cost, high authentication rate, and low authentication error rate.

\end{abstract}


\section{Introduction}
\emph {Confidentiality}, \emph{Integrity}, and \emph{Authentication} are the fundamental requirements for information security. Confidentiality ensures information is only available to unauthorized entities, integrity protects information accuracy and completeness during transmission, while authentication mainly assures the source of information. To provision those security functions, the typical approach is through upper-layer cryptographic algorithms/protocols, which usually provides computational security and might be comprised when adversaries have sufficient computation power.

As a complement, there is a flurry of research to provision information-theoretic security from the physical (PHY) layer \cite{Renna,Chen2013,chen2017,Wang2012,Liu2013,Zhang2015}. Information-theoretic security can ensure the aforementioned security attributes, even though adversaries have infinite computational capabilities. Based on the security objective, PHY-layer security can be roughly divided into two categories: PHY-layer secure transmission and PHY-layer message authentication, where the former targets confidentiality while the latter focuses on message integrity and sender authentication. In the literature, the two research areas are usually separately studied. In addition, there is extensive research on PHY-layer secure message transmission, aiming to improve the secrecy rate at which message can be securely delivered \cite{Subramanian2011,Mahdavifar2011,Thangaraj2007,Suresh2010,Hof2010,Maurer1,Csiszar1,Csiszar4,Wyner}. In contrast, the research on PHY-layer message authentication is inadequately studied, and needs further investigation.

\begin{figure}[!t]
\centering
\includegraphics[scale=0.35]{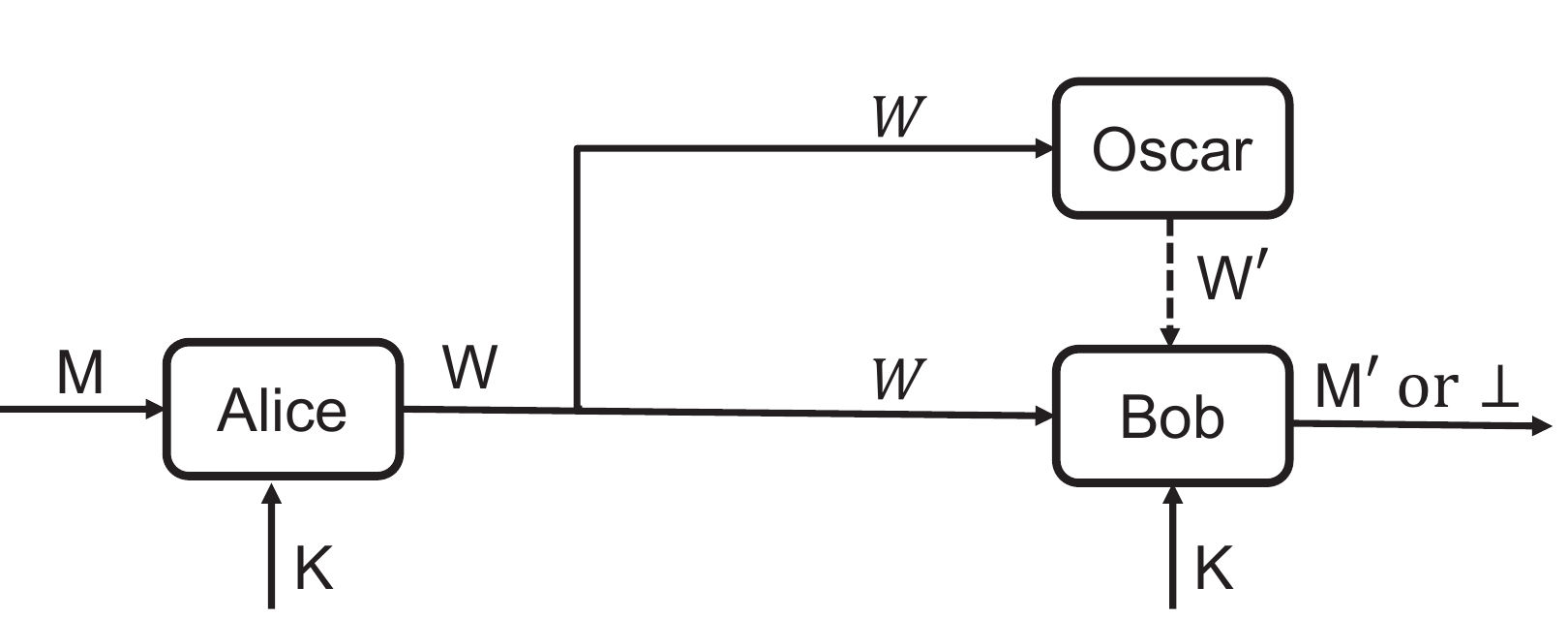}
\caption{The authentication model over noiseless channels.}
\label{noiselessauth}
\end{figure}

In the line of PHY-layer message authentication, the pioneering work by Simmons \cite{Simmons-Auth} proposes an authentication model over noiseless channels, as shown in Fig. \ref{noiselessauth}. Alice intends to transmit a message $M$ to Bob, while an adversary Oscar might launch two different types of man-in-the-middle attacks: 1) \emph {Impersonation attack}: forge the sender of the message; or 2) \emph {Substitution attack}: modify or replace the message. It is assumed that Alice and Bob share a common key $K$ in advance, which helps Bob identify the source of the message. The message $M$ and the key $K$ have distributions $P_M$ over message space $\mathcal{M}$ and $P_K$ over key space $\mathcal{K}$, respectively. Alice maps a pair $(M,K)$ to a codeword $W$, and sends $W$ over the noiseless channel. The adversary succeeds if Bob decodes the adversary's message and accepts it as a valid message from Alice. When performing multiple-message authentication, it is found that this model causes an entropy loss of the secret key. In fact, after $l$ times of authentication, the probability for successful attacks is at least $2^{H(K)/(l+1)}$, which quickly approaches 1 as $l$ increases \cite{Maurer}, where $H(\cdot)$ is the entropy function. 

\begin{figure}[!t]
\centering
\includegraphics[scale=0.38]{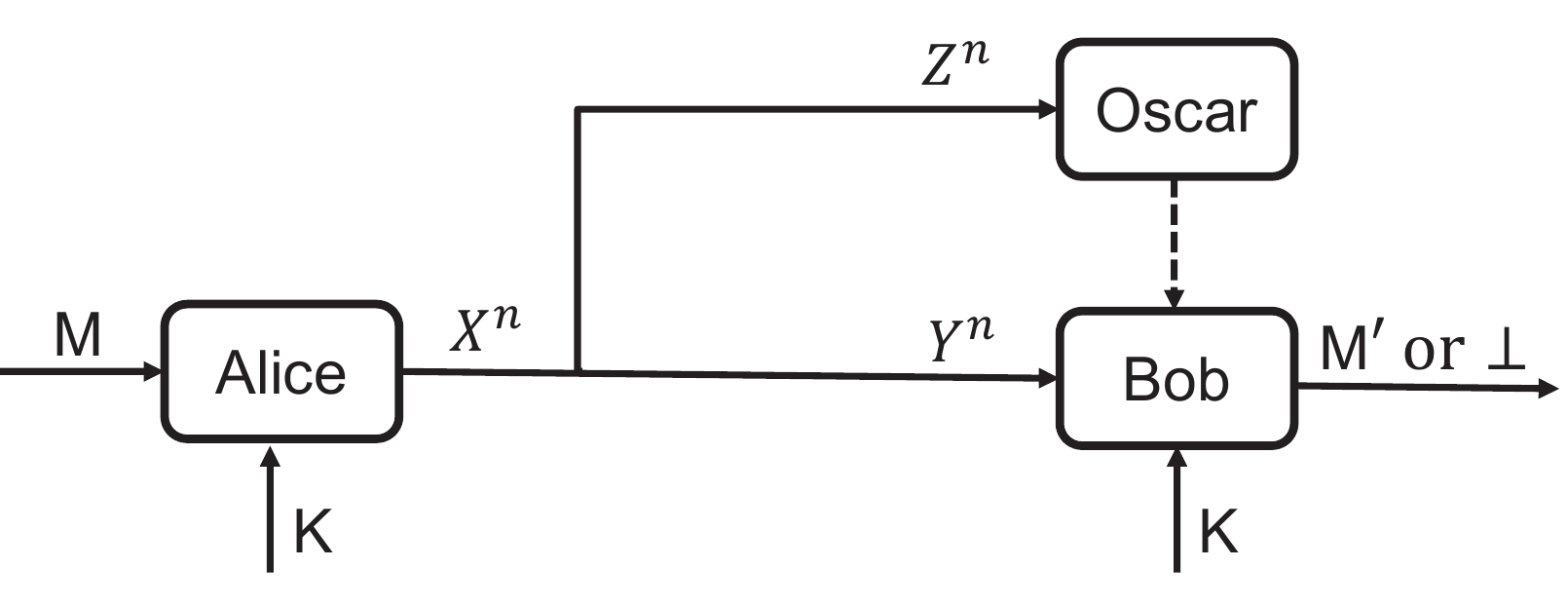}
\caption{The authentication model over noise channels.}
\label{Laimodel}
\end{figure}

In this paper, we aim to i) achieve information-theoretic security for multiple messages authentication with the same key; and ii) bridge the two separate areas of research in PHY-layer security. We propose a multi-message authentication framework over wiretap channel (as shown in Fig. \ref{Laimodel}), which integrates existing secure channel coding to achieve a high authentication rate. Specifically, Alice first encodes $M$ to $X^\nu$ using a channel coding method. Then, Alice generates a message tag $S$ of $M$ using a hash function
and employs a secure channel coding to encode $S$ to $X^n$. Finally, Alice transmits $(X^\nu, X^n)$ over wiretap channel $(W_1,W_2)$. Suppose that $<X^\nu, X^n>$ arrives at Bob as $<Y^\nu,Y^n>$, where Bob decodes $Y^\nu$ to $M'$ by the channel decoding function and decodes $Y^n$ to $S'$ by the secure channel code. Bob decides to reject or accept the authentication by checking the consistency of $<M',S'>$ (i.e., whether $S'$ is the tag of $M'$). To achieve information-theoretical security for a polynomial number of messages and attacks, we obtain a theorem (i.e., Theorem \ref{Th2}), which states the requirements/conditions for a authentication protocol to be information-theoretic secure to authenticate a polynomial number of messages. Furthermore, based on this theorem, we propose an authentication protocol with high efficacy. The authentication rate of the proposed authentication protocol approaches infinity when $n$ goes to infinity.

Based on theoretical results, we construct a feasible and efficient authentication protocol over binary symmetric wiretap channel (BSWC) by using Linear Feedback Shifting Register based (LFSR-based) hashing functions and strong secure polar codes.
Moreover, we evaluate the proposed protocol via extensive experiments.
The results demonstrate that, 1) by decreasing the secure rate, the strong secure coding scheme can provide the reliability of the main channel; and 2) the proposed authentication scheme has low time cost, high authentication rate, and low authentication error rate.

The main contributions of this work is summarized as follows:

\begin{itemize}
\item A multiple message authentication framework with the same secret key $K$ is proposed over wiretap channels. A theorem on the conditions for the authentication protocols to be information-theoretical secure is provided, with rigorous mathematical proof.
\item Based on obtained theorem, an authentication protocol is devised to achieve information-theoretic security with high efficiency. The authentication rate $\rho_{auth}=\rho_{tag}\cdot (C_s-\delta)$ for any fixed tag rate $\rho_{tag}$ and any small $\delta>0$. The authentication rate can approach infinity with $n$, where $n$ is the length of secure channel code $X^n$.
\item We bridge the gap between PHY-layer secure transmission and PHY-layer message authentication.
With the proposed framework and theorem, a strongly secure channel coding with exponentially small (in code length $n$) average probability of error can induce a multi-message authentication with information-theoretic security.
\item A feasible and efficient authentication protocol over BSWC is proposed by leveraging the lightweight LFSR-based hashing functions and secure polar codes. Extensive experiments validate the feasibility and efficiency of the proposed protocol.
\end{itemize}



The remainder of the paper is organized as follows.
The related work is reviewed in Section \ref{sec:related work}. Section \ref{sec:Preliminaries} introduces basic concepts and preliminaries that will be used in the subsequent sections. Section \ref{sec:model} presents the system model, including the authentication model, the adversary model, and security definitions. In Section \ref{sec:mutiauth}, a multiple-message authentication framework is proposed.
In Section \ref{sec:securityanalysis}, the theorem for authentication protocol to achieve information-theoretical security is provided
In Section \ref{sec:implementation}, we propose a multi-message authentication protocol and analyze its efficiency.
Section \ref{sec:casestudy} presents a efficient and feasible authentication protocol over BSWC.
In Section \ref{sec:performence}, we give the simulation studies for authentication over BSWC.
The concluding remarks are provided in Section \ref{sec:conclusion}.

\section{Related Work}\label{sec:related work}
\subsection{PHY-Layer Secure Transmission:}
The pioneering work on PHY-layer secure transmission in Wyner \cite{Wyner}, demonstrates that  information-theoretic security can be achieved, if the received signal at the attacker is a degraded version of that at the destination. This result is generalized by Csisz\'{a}r and K\"{o}ner \cite{Csiszar1}, in which the attacker's channel is not necessary to be a degraded version of the receiver's channel. Afterwards, secure transmission over noisy channels are extensively investigated in both theory and implementations \cite{Subramanian2011,Mahdavifar2011,Thangaraj2007,Suresh2010,Hof2010,Maurer1,Csiszar4,Renna,Chen2013,Wang2012,Lu2013,Liu2013,Zhang2015}.
Particularly, in \cite{Thangaraj2007}, a coset coding scheme by using the dual of low-density parity-check (LDPC) code is proposed to achieve weak secrecy on a binary erasure wiretap channel (BEWC). Extending this result, a coset coding scheme is proposed by leveraging the dual of short-cycle-free LDPC code to achieve the strong secrecy on a BEWC in \cite{Suresh2010}.
In \cite{Subramanian2011}, Subramanian \emph{et al.} propose a strongly secure channel coding scheme for binary erasure wiretap channel models (i.e., both the main cannel and the wiretapper's channel are binary erasure channels) by using large-girth  LDPC codes. In \cite{Hof2010}, polar codes is proposed as methods for approaching the secrecy capacity of general degraded and symmetric wiretap channels. In \cite{Mahdavifar2011}, Mahdavifar \emph{et al.} devise another channel coding algorithm based on polar codes for binary symmetric wiretap channel models (i.e., both the main cannel and the wiretapper's channel are binary symmetric channels).

\subsection{PHY-Layer Message Authentication:}
PHY-layer message authentication can be traced back to Simmons' work in \cite{Simmons-Auth}, where an authentication model over noiseless channels is proposed. In \cite{Maurer2}, the authentication is studied considering that the adversary has partial information regarding the key shared by the sender and receiver. Recently, authentication over noise channel models drawn increasing attentions \cite{KY+07,Barni,BLT12,FL+13,Jiang1,Jiang2}. The authentication over noise source model with a (noiseless) public discussion channel was studied by Korzhik \emph{et al.} in \cite{KY+07} and Barni in \cite{Barni}.
After that, the authentication over MIMO fading wiretap channels was considered by Baracca \emph{et al.} in \cite{BLT12} and Ferrante \emph{et al.} in \cite{FL+13}.
More recently, Jiang considered the keyless authentication problem in a noise channel model in \cite{Jiang1,Jiang2}.
The other related works also includes \cite{Liu2016,Lu2016,Ren2016}. In \cite{Liu2016}, a physical layer authentication mechanism was proposed by using the multipath effect between the sender and the receiver. In \cite{Ren2016}, a wireless physical-layer identification protocol by utilizing the unique features of the physical waveforms of wireless signals was presented. To achieve information-theoretic security for multiple messages authentication, Lai et al. \cite{Lai} study the message authentication over noisy channel, as shown in Fig. \ref{Laimodel}, where the channels from Alice to Bob and from Alice to Oscar can be regarded as a\emph{ wiretap channel model} (please refer to Section \ref{subsec:WCmodel}).
However, the authentication efficiency is bounded by the capacity of the channel from Alice to the adversary, denoted as ${I(X;Z)}$.

Different existing works, this work focuses on multi-massage authentication over wiretap channels with the same key. We integrate strongly secure channel coding in the authentication framework to achieve information-theoretical security. The conditions for the authentication protocol to be secure are obtained. With the proposed framework, we bridge the gap between PHY-layer secure transmission and PHY-layer message authentication. In this way, any advances of the area of (computationally efficient) secure channel coding will result in the improvement of message authentication. Moreover, we propose an authentication protocol which can satisfy the security requirements, with the efficiency of $\rho_{tag}\cdot (I(X;Y)-I(X;Z)-\delta)$ which approaches infinity with $n$.

\section{Preliminaries}\label{sec:Preliminaries}

\textbf{{Notations:}} Random variables are denoted by upper case letters (e.g., $X, Y, Z,\cdots$), their realizations are denoted by lower case letters (e.g., $x, y, z,\cdots$), and the domain of a random variable is denoted by calligraphic letters (e.g.,
$\cal{X}, \cal{Y}, \cal{Z}, \cdots$). Probabilities $P(X=x)$ and $P(X=x|Y=y)$ are denoted by $P_X(x)$ and $P_{X|Y}(x|y)$, or
$P(x)$ and $P(x|y)$, respectively. The following information theory terms can be found in
existing information theory books (e.g. \cite{CT91,CC2011}).

\begin{itemize}
\item  $I(X; Y)= \sum_{x, y}P(x, y)\log \frac{P(x, y)}{P(x)P(y)}$ is the \emph{mutual information} between $X$ and $Y$. $H(X)=-\sum_{x}P(x)\log P(x)$ and $H(X|Y)=-\sum_{x, y}P(x, y)\log P(x|y)$ is the
\emph{entropy} function and \emph{conditional entropy} function, respectively.
\item $x^n$ denotes a sequence of $x_1, \cdots, x_n.$
\item The \emph{type} of a sequence $x^n\in{\mathcal{X}^n}$ is the distribution $P_{x^n}$ on
$\mathcal{X}$ defined by
$P_{x^n}(a)=\frac{1}{n}N(a|x^n)$ for every $a\in \mathcal{X}$, where  $N(a|x^n)$ is the number of
occurrences of $a\in{\mathcal{X}}$ in $x^n$.
\item For any type  $P$ of length $n$ on ${\cal X}$, the set of sequences in $\mathcal {X}^n$ with  type $P$ is called a \emph{type class}  and is denoted by $T_P^n$.
\item \emph{Distance} between  random variables $X$ and $X'$ over ${\cal X}$ is  $\emph{SD}(X; X')=\sum_{x\in {\cal X}} |P_X(x)-P_{X'}(x)|.$
\item
\emph{Conditional
distance between  $X$ and $X$ given $Y$} is defined as
 \begin{equation}
{\tt SD}(X|Y; X)=\sum_{y\in {\cal Y}} P(y) \sum_{x\in {\cal X}} |P(x|y)-P(x) |.
\end{equation}\label{distance}
\item Function $negl(n)$ is negligible in $n$ if for any polynomial
$poly(n)$, $lim_{n\rightarrow\infty} {negl(n)}{poly(n)}= 0$.
\end{itemize}

\subsection{Universal Hash Functions}

Any function $f: A\rightarrow B$ with $|A|>|B|$ is called a \emph{hash function}.
A universal hash function is a hash function such that the output frequency occurs almost uniformly \cite{car,weg,Stinson,Stinson94}.
We now give a definition of the family of almost universal hash functions as follows.

\begin{definition} Let  ${\cal M}$  and ${\cal S}$ be two finite sets.
For  $\varepsilon> 0$, a collection of functions $\Psi$ from $\mathcal{M}$ to $\mathcal{S}$ is called
\emph{\em $\varepsilon$-almost weak universal ($\varepsilon$-AWU$_2$)} if
\begin{eqnarray}\label{eq: au0}
\forall (m,s)\in\mathcal{M}\times\mathcal{S},~~ \Pr[\psi: \psi(m)=s]\le \varepsilon;\\
\label{eq: au}
\forall m_1, m_2(\neq m_1)\in\mathcal{M},~~ \Pr[\psi: \psi(m_1)=\psi(m_2)]\le \varepsilon.
\end{eqnarray}
\end{definition}

The family of hash functions is $\varepsilon$-almost universal ($\varepsilon$-AU$_2$)  if the first condition is replaced by
  \begin{equation}\label{eq: au1}
  \forall (m,s)\in\mathcal{M}\times\mathcal{S},~~ \Pr[\psi: \psi(m)=s]=1/|\mathcal{S}|,
 \end{equation}
 i.e., all hash values are equally likely.

\subsection{Discrete Memoryless Channel}
A discrete channel with input alphabet $\mathcal{X}$ and output alphabet $\mathcal{Y}$ is defined as a
stochastic matrix  $W=\{W(y|x): x\in \mathcal{X}, y\in \mathcal{Y}\}$, where $W(\cdot |x)$ is the distribution of the channel output $Y$ given the input $X=x$, i.e.,
$W(y|x)=P_{Y|X}(y|x)$. In this case, we usually say $X$ and $Y$ are \emph{connected by channel} $W$. In this paper, we only consider a \emph{discrete memoryless channel} (DMC): suppose that the input sequence is $x^n=x_1,\cdots, x_n$ and the output sequence is $y^n=y_1, \cdots, y_n$, then $P_{Y^n|X^n}(y^n|x^n)=
\prod_{i=1}^n P_{Y_i|X_i}(y_i|x_i)=\prod_{i=1}^n W(y_i|x_i).$
For simplicity, we  denote $\prod_{i=1}^n W(y_i|x_i)$ by $W(y^n|x^n)$.

Suppose Alice wants to send messages to Bob over DMC $W$. Let her message domain be ${\cal S}$. Then the communication is described through a pair of mappings (called a \emph{coding scheme})  $(f, g),$ where
$f: {\cal S}\rightarrow {\cal X}^n$ and $g: {\cal Y}^n\rightarrow {\cal S}\cup \{\perp\}.$  When Alice wants to send $s\in {\cal S}$, he sends $f(s)$ through channel $W$. When Bob receives vector  $y^n$, he decodes the message as $s'=g(y^n)$, where $\perp$ denotes the detector of an error. Event $s\ne s'$ is called a \emph{decoding error}.  The set ${\cal C}=f({\cal S})$ is called the \emph{code book} of this coding scheme; $c=f(s)$ is called a  \emph{codeword}.

\subsection{Basics of  Wiretap Channel}\label{subsec:WCmodel}
Wiretap channel is first introduced by Wyner \cite{Wyner} and extended by Csiszar and Korner \cite{Csiszar1}.  A \emph{wiretap channel}  is defined by two DMCs $W_{1}:\mathcal{X}\rightarrow{\mathcal{Y}}$ and  $W_{2}:\mathcal{X}\rightarrow{\mathcal{Z}}$,
where $\mathcal{X}$ is the input alphabet from the sender Alice, $\mathcal{Y}$ is the output alphabet at the legitimate receiver Bob, and $\mathcal{Z}$ is the output alphabet at the wiretapper Oscar.
Alice aims to send private messages to Bob against Oscar.
Denote the domain of message of Alice by ${\cal S}$.
To send $s\in {\cal S}$, Alice sends it through $W_1$ with a coding scheme $(f, g)$.
Specifically, she first sends $x^n=f(s)$ into DMCs $W_1$ and $W_2$, from which Bob receives $y^n\in {\cal Y}^n$ and
Oscar receives $z^n\in {\cal Z}^n.$
Bob decodes $s'=g(y^n)\in {\cal S}\cup \{\perp\}.$
Let $X^n, Y^n, Z^n, S, S'$ be random variables for $x^n, y^n, z^n, s, s'$ respectively.
$\frac{1}{n}\log |{\cal S}|$ is the \emph{transmission rate}.

The goal of Alice and Bob is to maximize the transmission rate while keeping Oscar from knowing anything regarding $S$.
The security is defined as follows.

\begin{definition}
The sequence of coding schemes $\{(f_n, g_n)\}_n$ is called strongly secure channel coding
 with exponentially small (in $n$) average probability of error for  the wiretap channel  $(W_1, W_2)$ (denoted as {\bf{strongly secure channel coding}} for short), if there exists a constant $c>0$ such that the following conditions are satisfied:
\begin{eqnarray}\label{eq1}
\text{Reliability Condition:}~~~~~~~ \Pr{\{{S'}\neq{S}\}}\leq \exp({-c n})\\
\label{eq2}
\text{Strong Security Condition:}~~~~~{{I(S;Z^n)}}\leq \exp({-c n}).
\end{eqnarray}
For $R>0$, if ${{\frac{1}{n}}{\log|\mathcal{S}|}}\geq{R}$,
we called rate $R$ is \emph{\em securely achievable} for $(W_1, W_2)$.
The supremum  of securely achievable rates is called \emph{\em secret capacity} of the  wiretap channel and is denoted by  $C_{s}$.

\end{definition}

\section{System Model}\label{sec:model}
In this section, the authentication model is first presented, and the adversary model and the definition of secure authentication protocol are then elaborated.

\subsection{Authentication Model}
\begin{figure}[!t]
\centering
\includegraphics[scale=0.35]{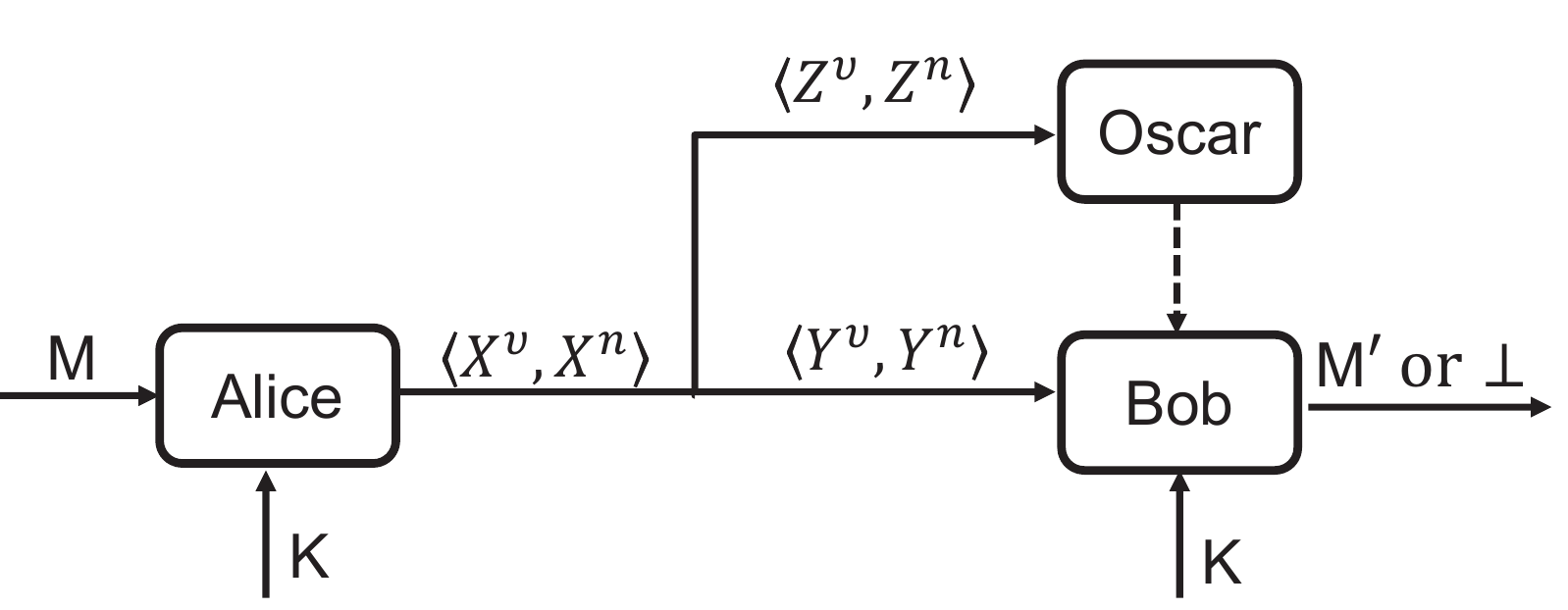}
\caption{The authentication model.}
\label{noiseauth}
\end{figure}

We consider a wiretap channel $W_{1}:\mathcal{X}\rightarrow {\mathcal{Y}}$, $W_{2}:\mathcal{X}\rightarrow {\mathcal{Z}}$.
Alice and Bob share a secret $K$ that is uniformly in a set ${\cal K}.$ They are connected by channel $W_1$. When Alice sends $X\in {\cal X}$, Bob and Oscar will receive $Y\in {\cal Y}$ and $Z\in {\cal Z}$, respectively.
Moreover, there is a noiseless channel form Oscar to Bob.
It is clear that the noiseless channel actually gives Oscar an advantage, since any noisy channel can be simulated with this noiseless channel by simply randomizing the transmitted signal. Note that, in wireless communications, the noiseless channel and noisy channel are the same wireless medium, where the former employs a Shannon channel code. Let ${\cal M}$ be a set of messages. As shown in Fig. \ref{noiseauth}, when Alice attempts to send  $M\in {\cal M}$ to Bob, they perform the following procedure.
\begin{itemize}
\item  Alice encodes $M$ into $X^\iota$ with a Shannon channel code for channel $W_1$, and then transmits it to Bob over channel $W_1$, and then, encodes $(M, K)$ into $X^n\in {\cal X}^n$ as an authentic with an encoder $f$ and sends it over wiretap channel $(W_1, W_2)$.
 \item Bob receives $Y^\nu$ and $Y^n$ from channel $W_1$. He then decodes $M'$ from $Y^\nu$ with Shannon channel code and decodes $D\in\{\top,\perp\}$ from $M'$, $Y^n$ and $K$ using a decoder (or decider) $g$, where $D=\top$ means that Bob accept $M'$ and $D=\perp$ means that Bob rejects $M'$.
\end{itemize}
Here, when $D=\top$, Alice indeed sends $M$ and $M=M'$, i.e., the decoded message $M'$ indeed is authenticated from Alice.

\begin{figure}[!t]
\centering
\includegraphics[scale=0.35]{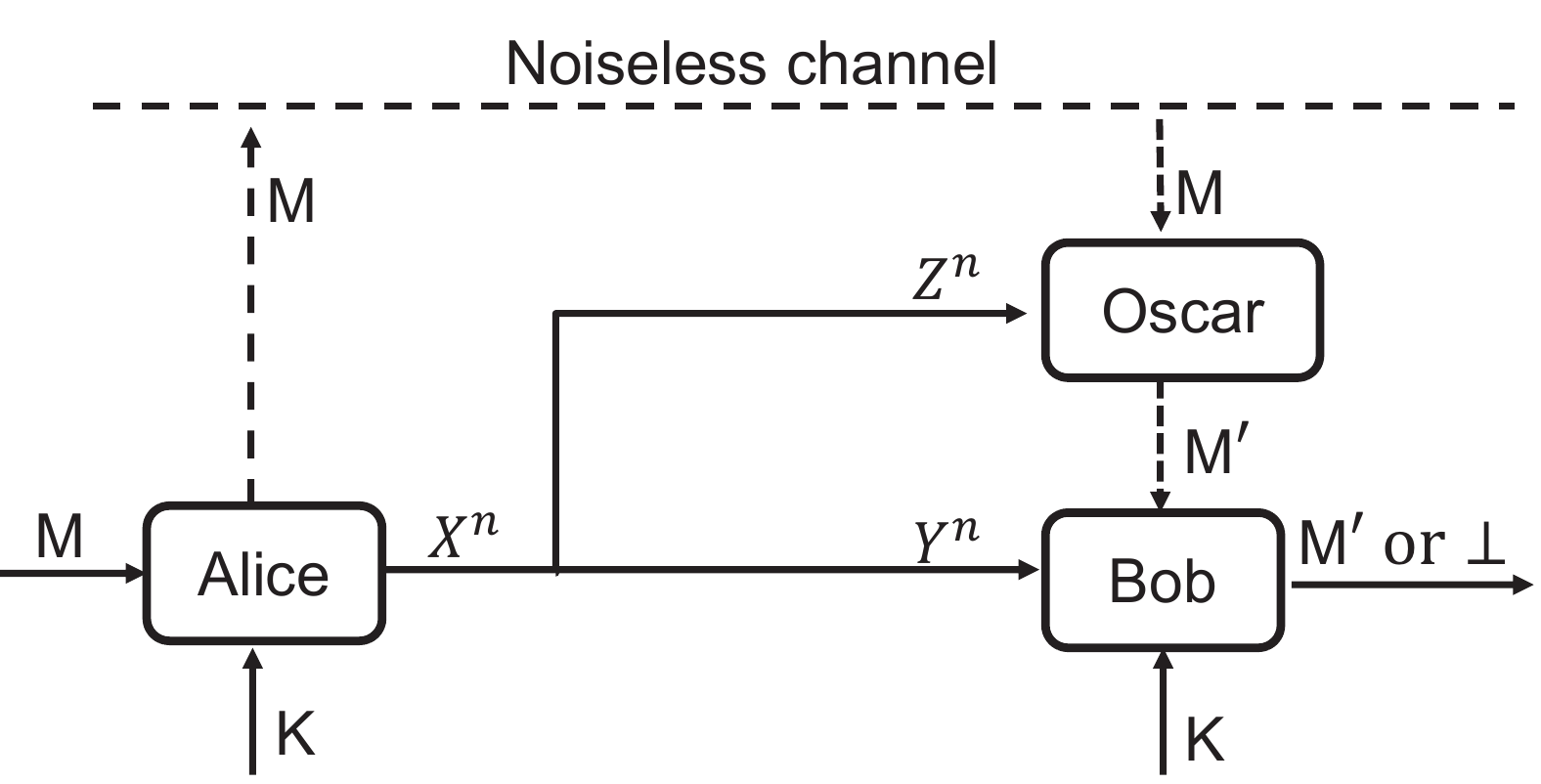}
\caption{The conceptual authentication model.}
\label{noiseauthnew}
\end{figure}

Without loss of generality, we can regard the information transmission with Shannon channel code as the information transmission over a noiseless channel. Thus, we consider that there exists a noiseless channel from Alice to Bob which is full controlled by Oscar. For ease of presentation, we adopt a conceptual authentication framework in the following, as shown in Fig. \ref{noiseauthnew}, which is equivalent to Fig. \ref{noiseauth}. Then, the authentication is performed as follows:
\begin{itemize}
\item Alice sends $M$ over  a public but unauthenticated  noiseless channel, and then, encodes $(M, K)$ into $X^n\in {\cal X}^n$ as an authentic with an encoder $f$ and sends over wiretap channel $(W_1, W_2)$.
 \item Bob receives $M'$ from the output of noiseless channel and  $Y^n$ from the output of channel $W_1$. He then decodes $D\in\{\top,\perp\}$ from $M'$, $Y^n$ and $K$ using a decoder (or decider) $g$, where $D=\top$ means that Bob receives $M'$ and $D=\perp$ means that Bob rejects $M'$.
\end{itemize}

\subsection {Adversary Model}

The communication link between Alice and Oscar is characterized by Channel $W_2$ and the link between Oscar and Bob is {\em noiseless}. Oscar's attack capability is formalized as follows.
\begin{itemize}
\item[1.] He can adaptively request Alice to authenticate any message $M$ of his choice. As a result, Alice normally authenticates $M$ to Bob through channel $(W_1, W_2)$ and Oscar receives $Z^n$ from Channel $W_2.$

\item[2.] He can adaptively sends any message $M'\in{\mathcal{M}}$ and vector ${Y'}^n\in {\cal Y}^n$ to Bob. Bob then decodes $M'$, ${Y'}^n$ and $K$ into $D\in\{\top,\perp\}$.
\end{itemize}

Oscar succeeds (denoted by \emph{Succ}) if $D=\top$ in item (2) occurs at least once.

\subsection {Security Definition}
The security concern consists of completeness and authentication.
Completeness essentially assures that when Oscar does not present, Bob should receive $M$ correctly with a high probability.
Authentication assures that the authentication failure event \emph{Succ} occurs negligibly.
Formally, we summarize it as follows.

\begin{definition} \label{def:auth}
 A cryptographic protocol  $\Pi$ for a wiretap channel $W_1: {\cal X}\rightarrow {\cal Y}, W_2: {\cal X}\rightarrow {\cal Z}$ is  a \emph{\em secure authentication protocol} if the following holds:
\begin{itemize}
\item[1.] {\bf Completeness. } \quad When the wiretapper Oscar does not present, there exists $\alpha>0$ such that $\Pr(D=\bot)\le \exp(-n\alpha),$ where $n$ is the number of use of the wiretap channel $(W_1,W_2)$ in the protocol.
\item[2.] {\bf Authentication. } For any wiretapper Oscar, the probability of success  $\Pr(\emph{\em Succ}(\mbox{Oscar}))$ is negligible in $n$.
\end{itemize}

If we only require  authentication to hold against Oscar that issues at most $t$ authentication queries  at item (1) of the adversary model,
then $\Pi$ is \emph{\em $t$-secure authentication} protocol.
\end{definition}
In our previous work \cite{Chen2015}, we define the efficiency
metric for a secure authentication protocol as follows.
\begin{itemize}
  \item [3. ]{\bf Efficiency.} The authentication rate is defined as $\rho_{auth} =\frac{1}{n} log |M|$, which is the ratio of the source message
length to the codeword length.
\end{itemize}

\section{Authentication of multiple message}\label{sec:mutiauth}
In this section, we first propose an authentication framework. Then, based on this framework, we conduct the security analysis to find the conditions/requirements for the authentication protocol to be secure (in Theorem 2)

\subsection{Authentication Framework}\label{scheme}

Let  $\{(f_n, g_n)\}_n$ be a secure channel coding for wiretap channel $W_1: {\cal X}\rightarrow {\cal Y}, W_2: {\cal X}\rightarrow {\cal Z}$; ${\cal S}=\{1, \cdots, 2^{n R_n}\}$ be the source messages for $(f_n, g_n)$; and $R_n$ be the code rate of $(f_n, g_n)$.
Let $\Psi=\{\psi_k\}_{k\in {\cal K}}$ be a collection of hash functions from ${\cal M}$ to ${\cal S}$.
Alice and Bob share a secret key $k\in {\cal K}.$
When Alice intends to send message $m\leftarrow {\cal M}$ to Bob, they perform as follows.
\begin{itemize}
\item[1.] Alice computes $s=\psi_k(m)$ (which is called the \emph{message tag}), encodes $x^n=f_n(s)$, and then sends $m$ and $x^n$ over noiseless channel and channel $(W_1, W_2)$, respectively.  Suppose that Bob receives $m'$ and ${y'}^n$ and Oscar receives $m$ and $z^n$, respectively.
 \item[2.] Based on $m', {y'}^n$, Bob decodes $s'=g_n({y'}^n)$. If $s'=\perp$ or  $\psi_k(m')\ne s'$, Bob rejects the message $m'$; otherwise accepts $m'$.
\end{itemize}

Next, we employ the proposed framework to authenticate a sequence $J$ of messages $M_1,M_2,\cdots,M_J$ using the same key $K$.
In such a scenario, the attacker can choose a time slot $j$ in which to initiate either an impersonation attack or a substitution attack.

For an impersonation attack at slot $j$, Oscar sends a message to the receiver
before the source sends anything. Oscar select the transmitted message is based on the information collected through the last $j-1$ rounds of authentication. The attacker is successful if Oscar's message is accepted as authentic at Bob.
For a substitution attack at slot $j$, Oscar intercepts the Alice's $j$th packet, modifies it, and sends the modified packet to the receiver. Oscar can make the modification using the information gathered in the past transmissions. The attacker is successful if the modified signal is accepted as authentic and the message is decoded incorrectly.

For convenience, we denote the random vectors $(x_1,\cdots,x_n)$ by $\vec{x}$ .
In slot $j$, if the intended message is $m_j$, Alice transmits it over the noiseless channel.
Then, Alice computes ${\vec{x}}_j=f_n(s_j)$ and transmits it over the wiretap channel, where $k$ is the key and message tag $s_j=\psi_k(m_j)$.
Though Oscar, Bob receives ${{m}_j'}$ and ${\vec{y}}_j'$. Oscar receives  ${m_j}$ and ${\vec{z}}_j$.
We denote the probability of successful impersonation
attack at the $j$th slot by $P_{I,j}$  and  the probability of successful
substitution attack at the $j$th  slot by $P_{S,j}$.

\subsection{Analysis on Attacks}
\emph{Impersonation attack:}
To initiate an impersonation attack in slot $j$, Oscar can use the information collected through $m_1,\vec{z}_1,\cdots,m_{j-1}, \vec{z}_{j-1}$.
Let $h_{j,im}$  be the strategy (or a function) employed by the source that maps $m_1,\vec{z}_1,\cdots,m_{j-1}, \vec{z}_{j-1}$ to $m_{oj}$, ${\vec{y}}_{o,j}$.
We also denote the decoded message tag by $s_{o,j}=g_n(\vec{y}_{o,j})$  and the message at the destination after Oscar's attack by $m_{oj}$.
For each $m_1,\vec{z}_1,\cdots,m_{j-1}, \vec{z}_{j-1}$, Oscar will adapt a strategy $h_{j,im}$ so that
the following probability is maximized:

{\small{\[\begin{split}
&P(s_{oj}=\psi_K(m_{oj})|m_1,{\vec{z}}_1,\cdots,m_{j-1}, {\vec{z}}_{j-1})\\
=&\sum_{k\in\mathcal{K}}{P(k|m_1,\vec{z}_1,\cdots,m_{j-1}, \vec{z}_{j-1})~{\theta(\psi_k(m_{oj}),s_{oj})}},
\end{split}\]}}

where
\[
\theta(r_1,r_2)=
\begin{cases}
1,&\text{if~} r_1=r_2;\\
0,&\text{otherwise,}\\
\end{cases}
~~~~~~\text{for any~} r_1,r_2\in{R}.\]

After receiving $j-1$ rounds of transmission, the probability of successful impersonation attack is
\begin{equation}\label{eq28}
\begin{aligned}
P_{I,j}=&\sum_{m_1,\vec{z}_1,\cdots,m_{j-1}, \vec{z}_{j-1}}P(m_1,{\vec{z}}_1,\cdots,m_{j-1}, {\vec{z}}_{j-1})\times\\
&\sup_{h_{j,im}}\biggl\lbrace{P(s_{oj}=\psi_K(m_{oj})|m_1,{\vec{z}}_1,\cdots,m_{j-1}, {\vec{z}}_{j-1})}\biggl\rbrace.
\end{aligned}
\end{equation}

\emph{Substitution attack:}
Oscar can also choose to invoke a substitution attack
after receiving the $j$th transmission i.e., it changes the content
of the $j$th package and sends it to the destination.
For a substitution attack, Oscar can adaptively interleave two types of attacks as follows.
In Type I attack, when  Alice sends $(m_j, \vec{x}_j)$, Oscar can revise $m$ to $m_{oj} (\ne m_j)$;
in Type II attack,  Oscar can send any  pair  $m_{oj} (\ne m_j), \vec{y}_{oj}$ to Bob noiselessly.
Oscar  succeeds,  if $s_{oj}=\psi_{k}(m_{oj})$
in Type I attacks (where $g_{n}(\vec{y}_{j})={s_{oj}}$),
 or $s_{oj}=\psi_{k}(m_{oj})$  in Type II attacks (where $g_{n}( \vec{y}_{oj})={s_{oj}}$).\\

{\bf Type I attack:} Oscar knows $m_1,\vec{z}_1,\cdots,m_{j}, \vec{z}_{j}$, and hance can choose $m_{oj}$ based on this information. Let $h_{1j,sb}$  be the strategy employed by the source that maps
$m_1,\vec{z}_1,\cdots,m_{j}, \vec{z}_{j}$ to $m_{oj}$. Then, Oscar transmits  $m_{oj}$  over the noiseless channel and does not modify the authentication information. Hence, $m_{oj}$ and $s_j$ are the message and
message tag at the destination after the attack. The attacker is successful if $s_j=\psi_k(m_{oj})$, where $s_j=g(\vec{y}_{j})$.
Obviously, for each observation $m_1,\vec{z}_1,\cdots,m_{j}, \vec{z}_{j}$, Oscar should choose $h_{1j,sb}$
such that

{\small{\[\begin{split}
&P(s_j=\psi_K(m_{oj})|m_1,{\vec{z}}_1,\cdots,m_{j}, {\vec{z}}_{j})~(1-\theta(m_{oj},m_{j}))\\
=&\sum_{k\in\mathcal{K}}{P(k|m_1,\vec{z}_1,\cdots,m_{j}, \vec{z}_{j})}\times\\
&~~~~~~~~~~~~~~~\biggl\lbrace{\theta(\psi_k(m_{oj}),\psi_k(m_j))}(1-\theta(m_{oj},m_{j}))\biggl\rbrace
\end{split}\]}}
is maximized.
Therefore, the probability of successful type I substitution attack after receiving $j$ rounds of transmission is

{\small{\begin{equation}\label{eq29}
\begin{aligned}
&P_{S_{1},j}=\sum_{m_1,{\vec{z}}_1,\cdots,m_{j}, {\vec{z}}_{j}}P(m_1,{\vec{z}}_1,\cdots,m_{j}, {\vec{z}}_{j})\times\\
&\sup_{h_{1j,sb}}\biggl\lbrace{P(s_j=\psi_K(m_{oj})|m_1,{\vec{z}}_1,\cdots,m_{j}, {\vec{z}}_{j})}~(1-\theta(m_{oj},m_{j}))\biggl\rbrace.
\end{aligned}
\end{equation}}}

\textbf{Type II attack:} Let $h_{2j,sb}$ be
the strategy employed by the source that maps $m_1,\vec{z}_1,\cdots,m_{j-1}, \vec{z}_{j-1}$ to $m_{oj}$, ${\vec{y}}_{o,j}$. After the opponent's attack, the decoded message tag is denoted as $s_{o,j}=g(\vec{y}_{o,j})$
and the message at the
destination is denoted as $m_{oj}$.
The attack is successful if $s_{oj}=\psi_k(m_{oj})$. For each
possible observation $m_1,{\vec{z}}_1,\cdots,m_{j}, {\vec{z}}_{j}$, the opponent will adopt a
strategy $h_{2j,sb}$ so that the following probability is maximized:

{\small{\[\begin{split}
&P(s_{oj}=\psi_K(m_{oj})|m_1,{\vec{z}}_1,\cdots,m_{j}, {\vec{z}}_{j})(1-\theta(m_{oj},m_{j}))\\
=&\sum_{k\in\mathcal{K}}{P(k|m_1,\vec{z}_1,\cdots,m_{j}, \vec{z}_{j})}\times\biggl\lbrace{{\theta(\psi_k(m_{oj}),s_{oj})}~(1-\theta(m_{oj},m_{j}))}\biggl\rbrace.
\end{split}\]}}

Hence, the probability for the $j$th type II substitution attack being successful
is

{\small{\begin{equation}\label{eq30}
\begin{aligned}
&P_{S_{2},j}=\sum_{m_1,{\vec{z}}_1,\cdots,m_{j}, {\vec{z}}_{j}}P(m_1,{\vec{z}}_1,\cdots,m_{j}, {\vec{z}}_{j})\times\\
&\sup_{h_{2j,sb}}\biggl\lbrace{P(s_{oj}=\psi_K(m_{oj})|m_1,{\vec{z}}_1,\cdots,m_{j}, {\vec{z}}_{j})}~(1-\theta(m_{oj},m_{j}))\biggl\rbrace.
\end{aligned}
\end{equation}}}

Then, the probability of successful
substitution attack after receiving $j$ rounds of transmission is
\begin{equation*}
P_{S,j}=\max\{P_{S_{1},j},P_{S_{2},j}\}.
\end{equation*}

\section{Security Analysis}\label{sec:securityanalysis}
Based on the proposed framework, we first analyze the information leakage about the key $K$ and message tag $S$, from an information theoretical point of view. The results for single-/multi-message authentication are given by Proposition 1 and Proposition 2, respectively. Then, based obtained results, we obtain a theorem (Theorem \ref{Th2}) which states the requirements/conditions for the authentication protocol to achieve information-theoretic security when authenticating a polynomial number of messages.

\subsection{Security Theorem}
When a strongly secure channel coding is employed in the authenticate framework, the attacker obtains no significant amount of information about secret key $K$ and the message tag $S$, in one time authentication, as given in Proposition 1.
\begin{proposition}\label{proposition1}
Let $M$, $S$ and $Z^n$ be the random variables defined in the authentication framework (Sec.V.A). Then, with a strongly secure channel coding, there exists a constant $\alpha>0$, such that
\begin{equation}\label{SZM}
I(S;Z^n|M)\leq{e^{-\alpha n}},
\end{equation}
\begin{equation}\label{KZM}
I(K;Z^n|M)\leq{e^{-\alpha n}}.
\end{equation}
\end{proposition}

\begin{IEEEproof}
 The conditional mutual information can be rewritten as
  \[\begin{split}
I(S;Z^n|M)&=H(Z^n|M)-H(Z^n|S,M)\\
&=^{(a)} H(Z^n|M)-H(Z^n|S)=I(Z^n;S)-I(Z^n;M)\\
&\leq{I(Z^n;S)}\leq^{(c)} {e^{-\alpha n}},\\
I(K;Z^n|M)&=I(K,M;Z^n)-I(M;Z^n)\\
&\leq^{(b)}{I(S;Z^n)-I(M;Z^n)}\\
&\leq{I(S;Z^n)}\leq^{(c)}{e^{-\alpha n}},~~~~~~~
\end{split}\]
where (a) and (b) are based on the fact that $M\rightarrow{MK}\rightarrow{S}\rightarrow{Z^n}$ forms a Markov chain, while (c) comes from  Definition 2.
\end{IEEEproof}

Next, we analyze the security for multiple-message authentication.

\begin{proposition}\label{proposition2}
When the same key $K$ is used to authenticate a sequence $j$ of messages $M_1,\cdots,M_j$, if a strongly secure channel coding and a $\varepsilon$-$AU_2$ hash function are employed, then there exists a constant $\alpha>0$, such that for any $b=1,\cdots,j$
\begin{eqnarray}
I(S_b;\vec{Z}_1,\cdots,\vec{Z}_j|M_1,\cdots,M_j)&\leq&{ e^{-\alpha n}}\\
I(K;\vec{Z}_1,\cdots,\vec{Z}_j|M_1,\cdots,M_j)&\leq&{j\cdot e^{-\alpha n}}
 \end{eqnarray}
where $S_j=\psi_K(M_j)$.
\end{proposition}

\begin{IEEEproof}
From inequality (\ref{SZM}), there exists a constant $\alpha>0$  such that $I(S_j;\vec{Z}_j|M_j)\leq{e^{-\alpha n}}$.
Given $(M_1,\cdots,M_j)=(m_1,\cdots,m_j)$, for any $b\in\{1,\cdots,j\}$, we have
\begin{eqnarray}
{\vec{Z}_b}\rightarrow{S_b}\rightarrow{K}\rightarrow{(S_1\cdots,S_{b-1})}\rightarrow{(\vec{Z}_1,\cdots,\vec{Z}_{b-1})}\\
{\vec{Z}_b}\rightarrow{S_b}\rightarrow{K}\rightarrow{(S_{b+1},\cdots,S_{j})}\rightarrow{(\vec{Z}_{b+1},\cdots,\vec{Z}_{j})}
 \end{eqnarray}
form two Markov chains.
Hence, given $m_1,\cdots,m_j$,
${\vec{Z}_b}\rightarrow{S_b}\rightarrow{\vec{Z}_1,\cdots,\vec{Z}_{b-1},\vec{Z}_{b+1},\cdots,\vec{Z}_{j}}$
forms a Markov chain.
Consequently,  from data processing inequality, the following holds
\begin{equation}
I(S_b;\vec{Z}_1,\cdots,\vec{Z}_j|m_1,\cdots,m_j)\leq{I(S_b,\vec{Z}_b|m_1,\cdots,m_j)}.
\end{equation}
Averaging over $m_1, \cdots, m_b$, we have
\begin{eqnarray}
I(S_b;\vec{Z}_1,\cdots,\vec{Z}_j|M_1,\cdots,M_j)\\
\leq{I(S_b,\vec{Z}_b|M_1,\cdots,M_j)}~~~~~~~~~\\
=I(S_b,\vec{Z}_b|M_b)\leq{e^{-\alpha n}}~~~~~~~~~~
\end{eqnarray}


Given $(M_1,\cdots,M_j)=(m_1,\cdots,m_j)$, we have $\vec{Z}_b \rightarrow K\rightarrow (\vec{Z}_1,\cdots,\vec{Z}_{b-1})$ forms a Markov chain for any $b\leq j$.
Hence, by data processing inequality, we have
$I(K;\vec{Z}_b|\vec{Z}_1,\cdots,\vec{Z}_{b-1},m_1,\cdots,m_j)\leq{I(K;\vec{Z}_b|m_1,\cdots,m_j)}$.
Averaging over $m_1, \cdots, m_b$, we have
\begin{equation*}
I(K;\vec{Z}_b|\vec{Z}_1,\cdots,\vec{Z}_{b-1},M_1,\cdots,M_j)\leq{I(K;\vec{Z}_b|M_1,\cdots,M_j)}.
\end{equation*}
Thus,
 \[\begin{split}
&I(K;\vec{Z}_1,\cdots,\vec{Z}_j|M_1,\cdots,M_j)\\
\leq&\sum^j_{b=1} I(K;\vec{Z}_b|\vec{Z}_1,\cdots,\vec{Z}_{b-1},M_1,\cdots,M_j)\\
\leq&\sum^j_{b=1} I(K;\vec{Z}_b|M_1,\cdots,M_j)\\
=&\sum^j_{b=1} I(K;\vec{Z}_b|M_b)\leq{j\cdot e^{-\alpha n}}.
 \end{split}\]
\end{IEEEproof}

Based on the above results, we have the following theorem regarding the requirements/conditions for a multi-message authentication protocol to be information-theoretical security.

\begin{theorem}
Suppose that $\{(f_n, g_n)\}_n$ is a strongly secure channel coding for wiretap channel $(W_1,W_2)$,
and $\Psi : \mathcal{M}\times \mathcal{K}\rightarrow\mathcal{S}$ is an $\varepsilon$-
AWU$_2$ hash function with $\varepsilon$ being negligible in $n$. Then, for any polynomial $t(\cdot)$ and sufficiently large $n$, the proposed protocol  $\Pi$ is $t(n)$-secure.\label{Th2}
\end{theorem}

From this theorem, we only have to construct a family of
hashing function and strongly secure channel code
satisfying the conditions above to achieve  security of the proposed
protocol. We will discuss how to design such kind of class of hash functions and
channel code to meet these requirements in Sec. \ref{sec:casestudy}.
The detailed proof will be provided in the next subsection.
\subsection{Proof of Theorem \ref{Th2}}
In what follows, we prove Theorem \ref{Th2}. The main idea is to prove that sender Alice can authenticate a polynomial number of messages using $K$, where the attacker Oscar can adaptively interleave polynomial number of the impersonation attacks or substitution attacks.

We now present some lemmas that will be used to prove Theorem \ref{Th2}.
The following lemma is from \cite[Lemma 1]{Csiszar2}, where $X$ has the form $f(X)$ for a function $f$ in \cite{Csiszar2}. Since $f$ is arbitrary except $|f({\cal X})|\ge 4$, the two lemmas are equivalent.
\begin{lemma}\label{lem-distance}
 Let $X$ and $Y$ be two random variables over ${\cal X}$ and ${\cal Y}$, respectively, where $|{\cal X}|\ge 4.$ Then
\begin{equation}
\frac{1}{2\ln 2}{\tt SD}(X|Y; X)^2\le I(X; Y)\le {\tt SD}(X|Y; X)\log \frac{|{\cal X}|}{{\tt SD}(X|Y; X)}.  \label{le: Ibound}
\end{equation}
\end{lemma}

\begin{lemma}
If the same
key $K$ is used to authenticate a sequence $J$ of messages $M_1,M_2,\cdots,M_j$ by using the proposed framework $j$ times.
Let
\begin{equation*}
\Omega_j=\{m_1,\cdots,m_j:I(K;\vec{Z}_1,\cdots,\vec{Z}_j|m_1,\cdots,m_j)>{(j\cdot e^{-\alpha n})}^{1/2}\}.
\end{equation*}
Then, we have
\begin{equation}
P(\Omega_j)\leq{(j \cdot e^{-\alpha n)^{1/2}}},
\end{equation}
Moreover, for all $(m_1,\cdots,m_j)\in{{\Omega}_j^c}$,
\begin{equation}
I(K;\vec{Z}_1,\cdots,\vec{Z}_j|m_1,\cdots,m_j)\leq{{(j\cdot e^{-\alpha n})}^{1/2}}.
\end{equation}
\end{lemma}
\begin{IEEEproof} By the definition of the conditional mutual information, we have
 \[\begin{split}
    &{(j\tau)}^{1/2}\sum_{\Omega_j} P(m_1,\cdots,m_j)\\
    \leq&{\sum_{\Omega_j} P(m_1,\cdots,m_j)I(K;\vec{Z}_1,\cdots,\vec{Z}_j|m_1,\cdots,m_j)}\\
    \leq&{I(K;\vec{Z}_1,\cdots,\vec{Z}_j|M_1,\cdots,M_j))}\leq{j\cdot e^{-\alpha n}}.
    \end{split}\]
Therefore,
$$P(\Omega_j)=\sum_{\Omega_j} P(m_1,\cdots,m_j)=\leq{(j\cdot e^{-\alpha n)^{1/2}}}.$$
It holds that $I(K;\vec{Z}_1,\cdots,\vec{Z}_j|m_1,\cdots,m_j)\leq{{(j\cdot e^{-\alpha n)}}^{1/2}}$ for all $(m_1,\cdots,m_j)\in{{\Omega_j^c}}$.
\end{IEEEproof}

For all $(m_1,\cdots,m_j)\in{{\Omega}_j^c}$, we define probability  $\widetilde{P}(k,{\vec{z}}_1,\cdots,{\vec{z}}_j)$ on
${\mathcal{K}}\times{{\vec{\mathcal{Z}}}^j}$
 by
\begin{eqnarray}
{\widetilde{P}}(k,{\vec{z}}_1,\cdots,{\vec{z}}_j)=\Pr(k,{\vec{z}}_1,\cdots,{\vec{z}}_j|m_1,\cdots,m_j).
 \end{eqnarray}
Denote $\hat{P}(k)$ and $\hat{Q}({\vec{z}}_1,\cdots,{\vec{z}}_j)$ as the marginal distribution of ${\widetilde{P}}$, respectively.

By Lemma \ref{lem-distance}, Proposition \ref{proposition2} and the definition of ${\Omega_j^c}$,  we can obtain that, for all $(m_1,\cdots,m_j)\in{{\Omega_j^c}}$,
\begin{equation}\label{eq:sd}
\begin{split}
&SD(K|\vec{Z}_1,\cdots,\vec{Z}_j, m_1,\cdots,m_j;K|m_1,\cdots,m_j)\\
\leq&\sqrt{2\ln 2 I(K;\vec{Z}_1,\cdots,\vec{Z}_j|m_1,\cdots,m_j)}\\
\leq&{\sqrt{2\ln 2}(j\cdot e^{-\alpha n})^{1/4}},
\end{split}
\end{equation}
where the conditional probability distance is denoted by
\begin{equation}\label{eq:sd1}
\begin{split}
&SD(K|\vec{Z}_1,\cdots,\vec{Z}_j, m_1,\cdots,m_j;K|m_1,\cdots,m_j)\\
=&\sum_{\vec{z}_1,\cdots,\vec{z}_j}\hat{Q}(\vec{z}_1,\cdots,\vec{z}_j)SD(K|\vec{z}_1,\cdots,\vec{z}_j, m_1,\cdots,m_j;K|m_1,\cdots,m_j).
\end{split}
\end{equation}

\begin{lemma}
For all $(m_1,\cdots,m_j)\in{{\Omega_j^c}}$, we define
\begin{equation}\label{eq:sd2}
\mathcal{O}_j=\{{\vec{z}}_1,\cdots,{\vec{z}}_j:d({\vec{z}}_1,\cdots,{\vec{z}}_j)>{(j\cdot e^{-\alpha n})^{1/8}}\},
\end{equation}
where
\begin{equation}
d({\vec{z}}_1,\cdots,{\vec{z}}_j)=SD(K|\vec{z}_1,\cdots,\vec{z}_j, m_1,\cdots,m_j;K|m_1,\cdots,m_j).
\end{equation}
Then,
\begin{equation}
\hat{Q}(\mathcal{O}_j)\leq{\sqrt{2\ln 2}(j\cdot e^{-\alpha n})^{1/8}}.
\end{equation}
\end{lemma}

\begin{IEEEproof} From Equation \ref{eq:sd}, \ref{eq:sd1} and \ref{eq:sd2}, we have
\begin{equation*}
\begin{split}
&{(j\cdot e^{-\alpha n})^{1/4}}\sum_{\mathcal{O}_j}{\hat{Q}}({\vec{z}}_1,\cdots,{\vec{z}}_j)\\
\leq&{\sum_{\mathcal{O}_j}{\hat{Q}}({\vec{z}}_1,\cdots,{\vec{z}}_j)d({\vec{z}}_1,\cdots,{\vec{z}}_j)}\\
\leq&{SD(K|\vec{Z}_1,\cdots,\vec{Z}_j, m_1,\cdots,m_j;K|m_1,\cdots,m_j)}\\
\leq&{\sqrt{2\ln 2}}(j\cdot e^{-\alpha n})^{1/4}
\end{split}
\end{equation*}
and thus
$${\hat{Q}}(\mathcal{O}_j)=\sum_{\mathcal{O}_j}{\hat{Q}}({\vec{z}}_1,\cdots,{\vec{z}}_j)\leq{\sqrt{2\ln 2}(j\cdot e^{-\alpha n})^{1/8}}.$$
\end{IEEEproof}

Based on the discussion above, we now present the details of  proof of Theorem \ref{Th2}.
\begin{IEEEproof}
From Lemma \ref{code}, the completeness requirement can be satisfied. Next, we will concentrate on the authentication property.\\

 \textbf{The Upper bound of $P_{I,j}$:}
We bound the success probability of impersonation attack at slot $j$.

 For $j\geq{2}$, ${(m_1,\cdots,m_{j-1})\in{{\Omega}_{j-1}^c}}$ and $({\vec{z}}_1,\cdots,{\vec{z}}_{j-1})\in{{\mathcal{O}}_{j-1}^{c}}$, we have
\begin{equation*}
\begin{split}
    &{P(r_{oj}=\psi_K(m_{oj})|{{\vec{z}}_1,m_1,\cdots,{\vec{z}}_{j-1},m_{j-1}})}\\
    &=\sum_k {P(k|{{\vec{z}}_1,m_1,\cdots,{\vec{z}}_{j-1},m_{j-1}})}~{{\theta(\psi_k(m_{oj}),r_{oj})}}\\
    &\leq\sum_k \biggl\lbrace{|P(k|{{\vec{z}}_1,m_1,\cdots,{\vec{z}}_{j-1},m_{j-1}})-P(k|{m_1,\cdots,m_{j-1}})|}\\
    &~~~~~~~~~~~~~~~~~~~~~~~+{P(k|{m_1,\cdots,m_{j-1}})}\biggl\rbrace~{{\theta(\psi_k(m_{oj}),r_{oj})}}\\
    &\leq {d({\vec{z}}_1,\cdots,{\vec{z}}_{j-1})}+\sum_k{P(k|{m_1,\cdots,m_{j-1}})}~{{\theta(\psi_k(m_{oj}),r_{oj})}}\\
\end{split}
\end{equation*}
\begin{equation*}
\begin{split}
        &\leq {d({\vec{z}}_1,\cdots,{\vec{z}}_{j-1})}+\sum_k{P(k)}~{{\theta(\psi_k(m_{oj}),r_{oj})}}\\
    &~~~~~~~~~~~~~~~~~~~~~~~~~~~~~~\text{($K$ and $M_1,...,M_j$ are independent)}\\
    &=(j\cdot e^{-\alpha n})^{1/8}+\frac{1}{|\mathcal{K}|}(|\mathcal{K}|\varepsilon)~~~\text{(by the definition of $\varepsilon$-AWU$_2$)}\\
    &=(j\cdot e^{-\alpha n})^{1/8}+\varepsilon
\end{split}
\end{equation*}

Then, by Equation (\ref{eq28}), the success probability of impersonation attack can be given as
\[\begin{split}
 P_{I,j}&=\sum_{m_1,\vec{z}_1,\cdots,m_{j-1}, \vec{z}_{j-1}}P(m_1,{\vec{z}}_1,\cdots,m_{j-1}, {\vec{z}}_{j-1})\times\\
        &~~~~~~~~~~~~~\sup_{h_{j,im}}\biggl\lbrace{P(s_{oj}=\psi_K(m_{oj})|m_1,{\vec{z}}_1,\cdots,m_{j-1}, {\vec{z}}_{j-1})}\biggl\rbrace\\
        &=\biggl\lbrace\sum_{\{\Omega_{j-1}^c\times{\mathcal{O}}_{j-1}^c\}^c}+\sum_{\Omega_{j-1}^c\times{\mathcal{O}}_{j-1}^c}\biggl\rbrace P(m_1,{\vec{z}}_1,\cdots,m_{j-1}, {\vec{z}}_{j-1})\times\\
        &~~~~~~~~~~~~~\sup_{h_{j,im}}\biggl\lbrace{P(s_{oj}=\psi_K(m_{oj})|m_1,{\vec{z}}_1,\cdots,m_{j-1}, {\vec{z}}_{j-1})}\biggl\rbrace\\
 \end{split}\]
\[\begin{split}
&\leq{3(j\cdot e^{-\alpha n})^{1/8}}+\sum_{\Omega_{j-1}^c\times{\mathcal{O}}_{j-1}^c}P(m_1,{\vec{z}}_1,\cdots,m_{j-1}, {\vec{z}}_{j-1})\times\\
        &~~~~~~~~~~~~~~~~~~~~~~~~~~~~~~~~~~~~~~~~~~~~~~~~~~\biggl\lbrace(j\cdot e^{-\alpha n})^{1/8}+\varepsilon\biggl\rbrace\\
        &\leq{\varepsilon+4(j\cdot e^{-\alpha n})^{1/8}}.
\end{split}\]

\textbf{The Upper bound of $P_{S,j}$:}
We first bound the success probability $P_{S_{1},j}$ of type I substitution attack in slot  $j$.

For $j\geq{2}$, ${(m_1,\cdots,m_{j})\in{{\Omega}_{j}^c}}$ and $({\vec{z}}_1,\cdots,{\vec{z}}_{j})\in{{\mathcal{O}}_{j}^{c}}$, we have
\begin{equation*}
\begin{aligned}
&P(s_{j}=\psi_K(m_{oj})|m_1,{\vec{z}}_1,\cdots,m_{j}, {\vec{z}}_{j}) ~(1-\theta(m_{oj},m_{j}))\\
=&\sum_{k\in\mathcal{K}}{P(k|m_1,\vec{z}_1,\cdots,m_{j}, \vec{z}_{j})}\times\\
&~~~~~~~~~~~~~~~~~~~~~~\biggl\lbrace{\theta(\psi_k(m_{oj}),\psi_k(m_j))}(1-\theta(m_{oj},m_{j}))\biggl\rbrace\\
\leq& {d({\vec{z}}_1,\cdots,{\vec{z}}_j)}+\sum_k{P(k|{m_1,\cdots,m_{j-1}})}\times\\
&~~~~~~~~~~~~~~~~~~~~~~\biggl\lbrace{\theta(\psi_k(m_{oj}),\psi_k(m_j))}(1-\theta(m_{oj},m_{j}))\biggl\rbrace\\
\leq& {d({\vec{z}}_1,\cdots,{\vec{z}}_j)}+\sum_k{P(k)}{\theta(\psi_k(m_{oj}),r_j)}(1-\theta(m_{oj},m_{j}))\\
&~~~~~~~~~~~~~~~~~~~~~~~~~~\text{($K$ and $M_1,...,M_j$ are independent)}\\
    =&(j\cdot e^{-\alpha n})^{1/8}+\frac{1}{|\mathcal{K}|}\cdot{\varepsilon{|\mathcal{K}|}}~~~\text{(by the definition of $\varepsilon$-AWU$_2$)}\\
    =&(j\cdot e^{-\alpha n})^{1/8}+\varepsilon
\end{aligned}
\end{equation*}

From Equation (\ref{eq29}), $P_{S_{1},j}$ can be rewritten as
\begin{equation*}
\begin{aligned}
&P_{S_{1},j}=\sum_{m_1,{\vec{z}}_1,\cdots,m_{j}, {\vec{z}}_{j}}P(m_1,{\vec{z}}_1,\cdots,m_{j}, {\vec{z}}_{j})\times\\
&~~~\sup_{h_{1j,sb}}\biggl\lbrace{P(s_j=\psi_K(m_{oj})|m_1,{\vec{z}}_1,\cdots,m_{j}, {\vec{z}}_{j})}~(1-\theta(m_{oj},m_{j}))\biggl\rbrace\\
&=\biggl\lbrace\sum_{\{\Omega_{j}^c\times{\mathcal{O}}_{j}^c\}^c}+\sum_{\Omega_{j}^c\times{\mathcal{O}}_{j}^c}\biggl\rbrace P(m_1,{\vec{z}}_1,\cdots,m_{j}, {\vec{z}}_{j})\times\\
&~~~~\sup_{h_{j,im}}\biggl\lbrace{P(s_{j}=\psi_K(m_{oj})|m_1,{\vec{z}}_1,\cdots,m_{j}, {\vec{z}}_{j})} ~(1-\theta(m_{oj},m_{j}))\biggl\rbrace\\
\end{aligned}
\end{equation*}
\begin{equation*}
\begin{aligned}
&\leq{3(j\cdot e^{-\alpha n})^{1/8}}+\sum_{\Omega_j^c\times{\mathcal{O}}_j^c}P(m_1,{\vec{z}}_1,\cdots,m_{j}, {\vec{z}}_{j})[(j\cdot e^{-\alpha n})^{1/8}+\varepsilon]\\
&\leq{\varepsilon+4(j\cdot e^{-\alpha n})^{1/8}}.
\end{aligned}
\end{equation*}

Then, we bound the success probability $P_{S_{2},j}$ of type II substitution attack in slot  $j$.
Similarly, we have
\begin{equation*}
\begin{aligned}
&{P(r_{oj}=\psi_k(m_{oj})|m_1,{\vec{z}}_1,\cdots,m_{j}, {\vec{z}}_{j})}(1-\theta(m_{oj},m_{j}))\\
    \leq&\varepsilon+(j\cdot e^{-\alpha n})^{1/8}
\end{aligned}
\end{equation*}

From Equation (\ref{eq30}) and following the same procedure in the proof of the upper bound of $P_{S_1,j}$, we have
    \[\begin{split}
P_{S_{2},j}=&\sum_{m_1,{\vec{z}}_1,\cdots,m_{j}, {\vec{z}}_{j}}P(m_1,{\vec{z}}_1,\cdots,m_{j}, {\vec{z}}_{j})\\
\sup_{h_{2j,sb}}&\biggl\lbrace{P(s_{oj}=\psi_k(m_{oj})|m_1,{\vec{z}}_1,\cdots,m_{j}, {\vec{z}}_{j})}(1-\theta(m_{oj},m_{j}))\biggl\rbrace\\
\leq&{\varepsilon+4(j\cdot e^{-\alpha n})^{1/8}}.
\end{split}\]
Thus, the success probability of Oscar after attacking $t(n)$ times from Oscar (and authenticating $t(n)$ messages with the same key) can be rewritten as
 \begin{equation*}
 \begin{split}
     P_{D}&\leq\sum_{j=1}^{t(n)}\max\{P_{I,j},P_{S_1,j},P_{S_2,j}\}\\
     &\leq{\sum_{j=1}^{t(n)}{\varepsilon+4(j\cdot e^{-\alpha n})^{1/8}}}~~~~~~~~~\text{(by $\varepsilon\geq\frac{1}{\mathcal{S}}$)}\\
      &\leq{{t(n)}\cdot\varepsilon+4\cdot t^2(n)\cdot e^{-\frac{1}{8}\alpha n}}
     \end{split}
 \end{equation*}
This is negligible as $t(n)$ is a polynomial and  $\varepsilon$ is negligible.
\end{IEEEproof}

\section{Authentication Protocol and Efficiency}\label{sec:implementation}
In this section, we first prove the existence of the authentication protocol that can meet the security requirements. Then, we propose an authentication protocol and analyze the efficiency.
\subsection{Existence}
The following result states that there exists a $\varepsilon$-AWU$_2$ class of hash functions with $\varepsilon$ being negligible in $n$.
\begin{theorem}\label{th1}\cite{Stinson}
Let $q$ be a prime power and let $i\geq 1$
be an integer. Then, there exists hence $(i + 1)/q$-AU$_2$ class
of $q^{i+2}$ hash functions from $\mathcal{M}$ to $\mathcal{S}$, where $|\mathcal{M}| = q^{2^i}$
and $|\mathcal{S}|= q$.
\end{theorem}

A set with size $v$ can be enumerated with $\log v$ bits.
It is clear that a family of $\varepsilon$-AU$_2$ hash functions  is $\varepsilon$-AWU$_2$.
Therefore, the above theorem implies that there exists a family of $(i+1)/q$-AWU$_2$  hash function indexed by
 $(i+2)\log q$ bits which can compress a message with length $2^i\log q$ to a tag with length $\log q$.
In practice, $\varepsilon=(i+2)/q$ and index length $(i+2)\log q$ should be small and the message length $2^i\log q$ should be  large.
If $i=8$ and $q=2^{50}$, then the index length is 62 bytes, $\varepsilon\approx 2^{-47}$, the message length is $1.5$M bytes with the tag of $50$ bits.

In \cite{Csiszar2}, Csiszar showed the following lemma by using random coding.

\begin{lemma}
{\em \cite{Csiszar2}}  Let $P$ be a type of length $n$ over ${\cal X}$ with $P(x)>0$ for all $x$ and $X$ is distributed according to $P$.
Consider   a wiretap channel $W_{1}:\mathcal{X}\rightarrow {\mathcal{Y}}$, $W_{2}:\mathcal{X}\rightarrow{\mathcal{Z}}$ with   $I(X;Y)>I(X;Z)+2\tau$ for some $\tau>0$.
Then, there exists a set    $\mathcal{C}_n\subseteq{{T}^n_{P}}$ with  size $2^{n(I(X;Y)-\tau)}$ and an equipartition $\phi: {\cal C}_n\rightarrow \{1, \cdots, \iota\}$ for  ${\cal C}_n$
with $\iota \leq{2^{n(I(X;Y)-I(X;Z)-2\tau)}}$
such that $\mathcal{C}_n$ is the code for channel $W_1$ with exponentially small (in $n$)
average probability of error and  $I(\phi(\tilde{X}^n);Z^n)$ is exponentially small (in $n$), where $\tilde{X}^n$ is uniformly random over ${\cal C}_n$.\label{code}
\end{lemma}

This lemma implies that a strongly secure channel coding can achieve secrecy rate of $\frac{1}{n}\log \iota$, for any $n$. Specifically, suppose that $(f_n, g_n)$ be the coding scheme for $W_1$, where  $f_n$ and $g_n$ are the encoding and decoding methods, respectively. For $s\in {\cal S}=\{1, \cdots, \iota\}$, $f_n(s)$ for the wiretap channel is to take $\tilde{x}^n\leftarrow \phi^{-1}(s)$, and decoding $g_n(y^n):=\phi[\bar{x}^n],$ for $\bar{x}^n=g_n(y^n).$, where $y^n$ is the receive message.

Let $\iota=2^u\leq{2^{n(I(X;Y)-I(X;Z)-2\tau)}}$.
Taking $q=2^u$ and $i=ploy(u)$ for some polynomial function $ploy(\cdot)$ (in Theorem \ref{th1}), we can conclude that the family of hash functions in this theorem satisfies the conditions in Theorem \ref{Th2}.
Based upon the discussion above, we have the theorem as follows.

\begin{theorem}
Let  $I(X;{Y})>{I(X;{Z})}+\tau$ for some constant $\tau>0$, where $Y$, $Z$ are
the outputs of wiretap channel $(W_1,W_2)$ with input $X$; and $P_X$ is a type $P$
with $P(x) > 0$.
Then, for any polynomial $t(\cdot)$ and sufficiently large $n$, there exists a $t(n)$-secure authentication protocol $\Pi$.\label{Th3}
\end{theorem}

\begin{IEEEproof}
It can be directly obtained from Theorem \ref{Th2}, Theorem \ref{th1}, and Lemma \ref{code}.
\end{IEEEproof}

In \cite{Wyner}, the concepts of partial ordering of DMC with common input alphabet were introduced.
The  single-letter characterization of the relation  \emph{channel} $W_1$ \emph{is more capable than channel} $W_2$ if $I(X;Y)\geq{I(X;Z)}$ for every input $X$.
The relation $channel\ W_1$ $is\ less\ noisy\ than$ $channel\ W_2$ was single-letter characterized by the property that for every Markov chain $U\rightarrow{X}\rightarrow{Y,Z}$, we have $I(U;Y)\geq{I(U;Z)}$. The relation \emph{channel} $W_2$ \emph{is no less noisy than channel} $W_1$ was characterized by the property that there exists a Markov chain $U\rightarrow{X}\rightarrow{Y,Z}$ such that $I(U;Y)>{I(U;Z)}$.
The following theorem show that the condition $I(X;{Y})>{I(X;{Z})}+\tau$ in Theorem \ref{Th3} can be loosened to the condition that
  $W_2$ is no less noisy than $W_1$.

\begin{corollary}
If $W_2$ is no less noise than $W_1$.
Then, for any polynomial $t(\cdot)$ and sufficiently large $n$, there exists a $t(n)$-secure authentication protocol $\Pi$ for wiretap channel $(W_1,W_2)$.\label{cor1}
\end{corollary}

\begin{IEEEproof}
It is directly obtained from Theorem \ref{Th3} and the definition that $W_2$ is no less noisy than $W_1$.
\end{IEEEproof}

\subsection{High-Efficiency Authentication Protocol}
Note that, in Lemma \ref{code}, $\iota$ can be any number no more than $2^{n(I(X;Y)-I(X;Z)-2\tau)}$ and $\tau$ is any positive number.
By \cite[Theorem 3]{Csiszar1}, if $W_1$ is more capable than $W_2$, then $$C_s=\max_{P_X} I(X;Y)-I(X;Z).$$
Thus, from Lemma \ref{code} and \cite[Theorem 3]{Csiszar1}, we have the following lemma.
\begin{lemma}\label{le: Cs}
There exists a sequence of types $\{P_n\}_n$ over ${\cal X}$ such that code rate $R_n$ for coding scheme $(f_n,g_n)$ satisfies ${\lim}_{n\rightarrow \infty} R_n=C_s$.
\end{lemma}

In order to obtain a authentication protocol, which satisfy the security requirements and is efficient, we use the strongly secure channel coding induced by Lemma \ref{code}.
Next, we only need to specify $\tau$, $\Psi$, $\mathcal{M}$, $\mathcal{K}$, $\mathcal{R}$ and $\varepsilon$.
For our construction, the constraint for $\mathcal{S}$ is
$\frac{\log {\mathcal S}}{n}<I(X;Y)-I(X;Z)+\tau ~\text{(by Lemma \ref{code})},$
 where $\tau$ only has the constraint
 $I(X;Y)>I(X;Z)+\tau ~
 \text{(by Lemma \ref{code})}.$
So for any  $\delta\in (0,I(X;Y)-I(X;Z))$, we can define $\tau=I(X;Y)-I(X;Z)-\delta/2$ and then set   $|\mathcal{S}|=2^{n(I(X;Y)-I(X;Z)-\delta)}$.      Then, we have
\begin{equation}
\rho_{chan}=\frac{\log |\mathcal{S}|}{n}=I(X;Y)-I(X;Z)-\delta
\end{equation}
 for any  $\delta\in (0,I(X;Y)-I(X;Z))$.

Let $\tau=I(X;Y)-I(X;Z)-\delta/2$. Then, $\rho_{auth}=[I(X;Y)-I(X;Z)-\delta]\cdot\rho_{tag}.$
Further, we realize $\varepsilon$-AWU$_2$ $\Psi$ with  $\frac{{\lambda(n)}+1}{q}$-AU$_2$ in Theorem \ref{th1},
 where $|\mathcal{S}|=q=2^{n(I(X;Y)-I(X;Z)-\delta)}$, $|\mathcal{K}|=q^{{\lambda(n)}+2}$, $|\mathcal{M}| = q^{2^{{\lambda(n)}}}$, $\varepsilon=\frac{{\lambda(n)}+1}{q}$.
It is easy to verify that under this setup,  the security condition in  our authentication theorem is satisfied as long as ${\lambda(n)}\leq 2^{{\omega n}}$ for some $\omega\in (0, \rho_{chan})$. As a result, $\rho_{tag}=2^{{\lambda(n)}}$ and
$\rho_{auth}=[I(X;Y)-I(X;Z)-\delta]2^{\lambda(n)}$, where ${\lambda(n)}
\leq 2^{{\omega}n}$ for some $\omega\in (0, \rho_{chan})$.
The details of the message authentication protocol are shown in Protocol \ref{EAM}.

\begin{algorithm}[hptb]\label{EAM}
\caption{High Efficiency Authentication Protocol}
\label{alg:sg}
\small{
\textbf{Preliminaries:}\\
\begin{itemize}
  \item Let $\triangle I=I(X;Y)-I(X;Z)$,  $\delta\in(0,\triangle I)$, $\tau=\triangle I-\delta/2$,
 $\omega\in (0, \triangle I-\delta)$, and $q=2^{n(\triangle I-\delta)}$.
  \item Let $(f_n,g_n)$ be the secrecy capacity achievable strong secrecy coding (in Lemma 3), where $n$ is the code length.
  \item Let $\Psi=\{\psi_k\}_{k\in {\cal K}}$ be a collection of $\varepsilon$-AWU$_2$ hash functions from ${\cal M}$ to ${\cal S}$, in which, $|\mathcal{M}| = q^{2^{{\lambda(n)}}}$, $|\mathcal{S}|=q$, $|\mathcal{K}|=q^{{\lambda(n)}+2}$,  $\varepsilon=\frac{{\lambda(n)}+1}{q}$, and
 $\lambda(n)\leq 2^{\omega n}$.
 \item Let $k\in {\cal K}$ be the secret key shared by Alice and Bob.
\end{itemize}

\textbf{Protocol:}

If Alice intends to send and authenticate message $m\in{\cal M}$ to Bob, then they perform  the following protocol.
\begin{itemize}
\item[1.] Alice first computes the message tag $s=\psi_k(m)$, and encodes $x^n=f_n(s)$ with the strongly secure channel coding.
          And then, Alice transmits $m$ and $x^n$ over noiseless channel and wiretap channel $(W_1, W_2)$, respectively.
 \item[2.] Based on the received information $m'$ and ${y'}^n$ from noiseless channel and wiretap channel $(W_1, W_2)$, respectively, Bob first decodes $s'=g_n({y'}^n)$. Then Bob verifies if $m'$ is sent from Alice as follows.
       If $s'=\perp$ or  $\psi_k(m')\ne s'$, Bob rejects the message $m'$; otherwise accepts $m'$.
\end{itemize}
}
\end{algorithm}

\subsection{Efficiency}
The authentication rate $\rho_{auth}$ can be rewritten as $\rho_{auth}=\rho_{tag}\cdot\rho_{chan}$,
 where $\rho_{tag}=\frac{\log |\mathcal{M}|}{\log |\mathcal{S}|}$  and  $\rho_{chan}=\frac{\log |\mathcal{S}|}{n}$.
We name $\rho_{tag}$ the  {\em tag rate} and  $\rho_{chan}$ the  {\em channel coding rate}, respectively.
Combining with Lemma \ref{le: Cs}, we can have the following theorem.

\begin{theorem} If $W_1$ is more capable than $W_2$, for any $\delta\in (0, C_s)$, taking  $\tau=C_s-\delta/2$,
 the proposed protocol is $t(n)$-secure (or polynomial secure) with
\begin{equation}
\rho_{auth}=(C_s-\delta)\cdot 2^{\lambda(n)},
\end{equation}
 where ${\lambda(n)}\leq 2^{{\omega}n}$ for some $\omega\in (0, \rho_{chan})$, and  $C_s=\max_{P_X} I(X;Y)-I(X;Z)$.
Furthermore,  if $\lim_{n\rightarrow{\infty}}{\lambda(n)}=\infty$, then
\begin{equation}
\lim_{n\rightarrow{\infty}}\rho_{auth}=\infty.
\end{equation}\label{Th4}
\end{theorem}

From Theorem \ref{Th4}, we have the following corollary.

\begin{corollary}
If $W_2$ is no less noisy than $W_1$, for any polynomial $t(\cdot)$ and a sufficiently large $n$, there exists a $t(n)$-secure authentication protocol $\Pi$ with
\begin{equation}
\rho_{auth}=(C_s-\delta)\cdot 2^{\lambda(n)},
\end{equation}
for wiretap channel $(W_1,W_2)$, where $C_s$ is the secrecy capacity of $(W_1,W_2)$ denoted as $C_s=\max_{U\rightarrow X\rightarrow YZ} I(U;Y)-I(U;Z)$.\label{cor2}
\end{corollary}

\begin{IEEEproof}
By Corollary 2 in \cite{Csiszar1}, we have  $C_s=\max_{U\rightarrow X\rightarrow YZ} I(U;Y)-I(U;Z)$.
Thus, there exists a RV $U$ such that $U\rightarrow X\rightarrow YZ$, and $I(U;Y)-I(U;Z)>C_s-\delta/4$ for any $\delta\in(0,C_s)$.
By using $U$ to replace $X$ in Theorem \ref{Th4} and taking $\tau=C_s-\delta/4$, it can be proved, based on Theorem \ref{Th4}.
\end{IEEEproof}

Note that, in \cite{Lai}, a capacity achieving codebook is divided
into $|\mathcal{K}|$ subsets, each of which is further partitioned into $|\mathcal{M}|$ bins, such that the information of the key
 $K$ can be hidden from the attacker.
Thus, the authentication rate of this protocol can be given as $\frac{1}{n}|\mathcal{M}|=I(X;Z)-\xi$ for a certain $\xi>0$.
The authentication rate of the proposed protocol can reach infinity when $n$ goes to infinity by employing the security channel code and $\varepsilon$-AU$_2$ hash functions.

\section{Implementation: authentication over BSWC}
\label{sec:casestudy}

In this section, we consider  message authentication problem over a binary symmetric wiretap channel (BSWC), where the main channel and the wiretapper's channel are the binary symmetric channel with crossover probability $p$ and $q$, respectively, denoted by BSC$(p)$ and BSC$(q)$.
From Theorem \ref{Th2}, we only need to design a $\varepsilon$-$AWU_2$ hash functions and a strong secure channel coding, which meet the requirements in this theorem, so as to achieve information-theoretic security.
To this end, we study how to meet these requirements by leveraging LFSR-based hash functions and strong secure polar code.

\subsection{$\varepsilon$-AWU$_2$ Hash Functions}
Let the message space $\mathcal{M}$ and tag space $\mathcal{S}$ be the set of binary strings of length $t$ and $u$, respectively.
We consider a specific hash  functions as follows.
A family of hash functions $\Psi: \mathcal{M} \rightarrow \mathcal{S}$ is $\oplus$-linear if and only if
\begin{equation}\label{xorhashing}
  \psi_k(m\oplus m')=\psi_k(m) \oplus \psi_k(m')
\end{equation}
 for any $m, m'\in\mathcal{M}$.

For any $m, m'\in\mathcal{M}$,  we have $\Pr[\psi: \psi(m)=\psi(m')]=\Pr[\psi: \psi(m)\oplus\psi(m')=\textbf{0}]=\Pr[\psi: \psi(m\oplus m')=\textbf{0}]$.
It further equals to $\Pr[\psi: \psi(m)=\textbf{0}]$ for any $m\in\mathcal{M}$, where $\textbf{0}$ is the  zero string.
Thus, a family of  $\oplus$-linear hash functions is $\varepsilon$-AWU$_2$ if and only if
\begin{equation}\label{eq: balanced}
\forall (m,s)\in\mathcal{M}\times\mathcal{S}, ~~~ \Pr[\psi: \psi(m)=s]\leq\varepsilon.
\end{equation}
The family of $\oplus$-linear hash functions satisfying Equation (\ref{eq: balanced}) is called $\varepsilon$-balanced according to Krawczuk's work in \cite{Krawczyk94}.

Carter and Wegman \cite{car} give a strong universal$_2$ family of hash functions
\begin{equation}
\left\{\psi_{A,b}:\psi_{A,b}(m)= m\cdot A+b\right\}_{A,b}
\end{equation}
where
$A$ is an $t\times u$ Boolean matrix, $m(\neq \textbf{0})$ is a message with length $t$, and $b$ is a binary vector with length $u$.
However, the key length is $u(t+1)$, which is too expensive for key distribution and storage.

To solve such a problem, Krawczuk in \cite{Krawczyk94} constructs a family of $\varepsilon$-AWU$_2$ hash functions by slightly modifying Carter and Wegman's method.
Specifically, Krawczuk provides an efficient construction of matrix $A$ with a \emph{Linear Feedback Shifting Register} (LFSR) to shorten the key length as follows.
Let $p(x)$ be an irreducible polynomial over $GF(2)$, and
$(a_0, a_1,\cdots,a_{u-1})$ be the initial state of a LFSR corresponding to the coefficients of $p(x)$.
If $a_0, a_1,\cdots$ is the bit sequence generated by the LFSR,  the matrix $A$ can be expressed as
\begin{equation}A=
\left(
  \begin{array}{cccc}
    a_0 & a_1 & \cdots & a_{u-1}\\
   a_1 & a_2 & \cdots & a_u \\
    \vdots& \vdots & \ddots & \vdots \\
    a_{t-1}& a_t & \cdots & a_{u+t-2} \\
  \end{array}
\right)
\end{equation}

Krawczuk \cite{Krawczyk94} shows that the LFSR-based construction defined above is $\varepsilon$-balanced (i.e., $\varepsilon$-AWU$_2$) for $\varepsilon\leq\frac{t}{2^{u-1}}$, and its key length is reduced from $t(u+1)$ to $3u$ (i.e., $u$ bits for $b$, $u$ bits for the generator polynomial $p(x)$, and $u$ bits for the initial state of the LFSR).

\subsection{Strong Secure Polar Codes}\label{polarcode}

\begin{figure}[!t]
\centering
\includegraphics[scale=0.4]{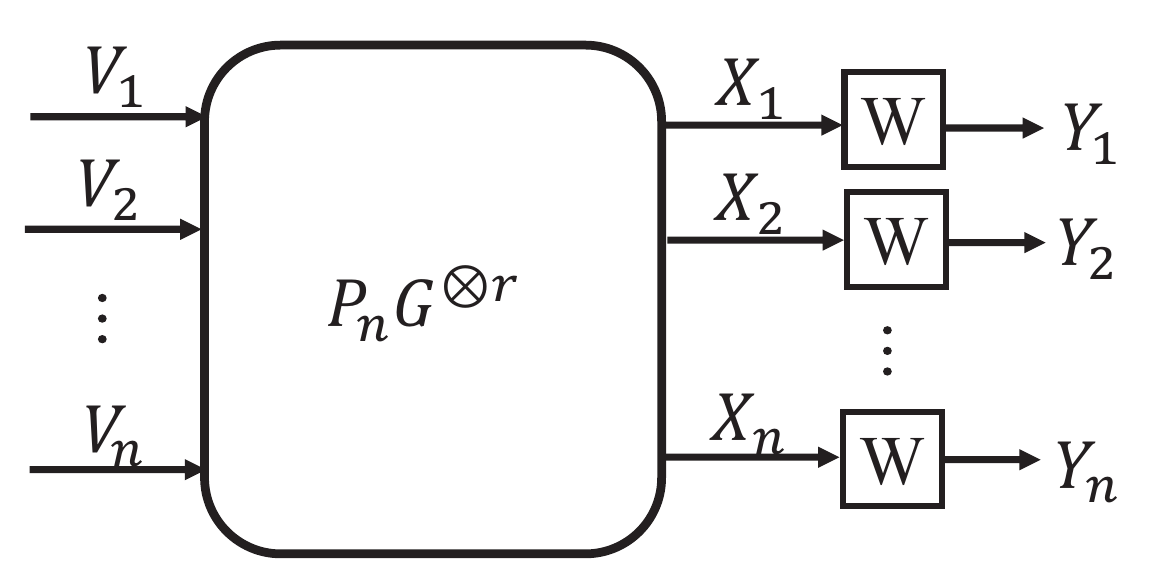}
\caption{The encoding process of polar codes.}
\label{PG}
\end{figure}

Polar code is introduced by Arikan in \cite{Arikan2009}, which can achieve the capacity of any binary-input symmetric DMCs with  low encoding and decoding complexity.
Let
\begin{equation}\label{eq: G}
  G=\left(
      \begin{array}{cc}
        1 & 0 \\
        1 & 1 \\
      \end{array}
    \right),
\end{equation}
and $G^{\otimes r}$ be the $m$-th Kronecker power of $G$.
For any ${V^n}\in\{0,1\}^n$, $V^n$ is encoded as $X^n=V^nP_nG^{\otimes r}$, where $n=2^r$ and $P_n$ is the bit-reversal permutation matrix with size $n\times n$. As shown in Fig. \ref{PG}, $X^n$ is sent over a binary-input channel symmetric DMC $W$ $n$ times independently.

By defining
\begin{equation}\label{tildeW}
  \widetilde{W}(y^n|v^n)=W^n(y^n|v^nP_nG^{\otimes r}),
\end{equation}
for any $i\leq n$, Arikan in \cite{Arikan2009} defines a channel $W_i:\{0,1\}\rightarrow \mathcal{Y}^n\times \{0,1\}^{i-1}$ as
 \begin{equation}\label{Wi}
{W}_i(y^n,v^{i-1}|v_i)=\frac{1}{2^{n-1}}\sum_{\textbf{e}\in\{0,1\}^{n-i}}\widetilde{W}(y^n|v^i,\textbf{e}).
\end{equation}
Arikan shows that as $r$ grows, it leads to channel polarization, i.e., for any $i$, $W_i$ approaches either a noiseless channel (i.e., good channel) or a pure-noise channel (i.e., bad channel).
Denoting Bhattacharyya parameter of channel $W$  as
\begin{equation}\label{eq: Bhattacharyya}
Z(W)=\sum_{y\in{\mathcal{Y}}}\sqrt{W(y|0)W(y|1)},
\end{equation}
the index sets of the good channels and bad channel can be defined as follows:
\begin{eqnarray}
  \mathcal{G}_n(W,\beta) &=& \left\{i\in[n]:Z(W_i)< 2^{-n^\beta}/n\right\} \\
  \mathcal{B}_n(W,\beta) &=& \left\{i\in[n]:Z(W_i)\geq 2^{-n^\beta}/n\right\}
\end{eqnarray}
where $\beta<1/2$ is a fixed constant, and $[n]=\{1,2,\cdots,n\}$.

Based on Arikan's work, a strong-security coding scheme with polar codes is proposed by Mahdavifar and Vardy in \cite{Mahdavifar2011}.
Let $C(W_i)$ be the capacity of channel $W_i$.
Define the index set of $\sigma_n$-poor bit-channel as
\begin{equation}\label{eq: poorset}
  \mathcal{P}_n(W,\sigma_n)=\left\{i\in[n]:C(W_i)\leq\sigma_n\right\}
\end{equation}
for some positive constant $\sigma_n$.

\begin{figure}[!t]
\centering
\includegraphics[scale=0.342]{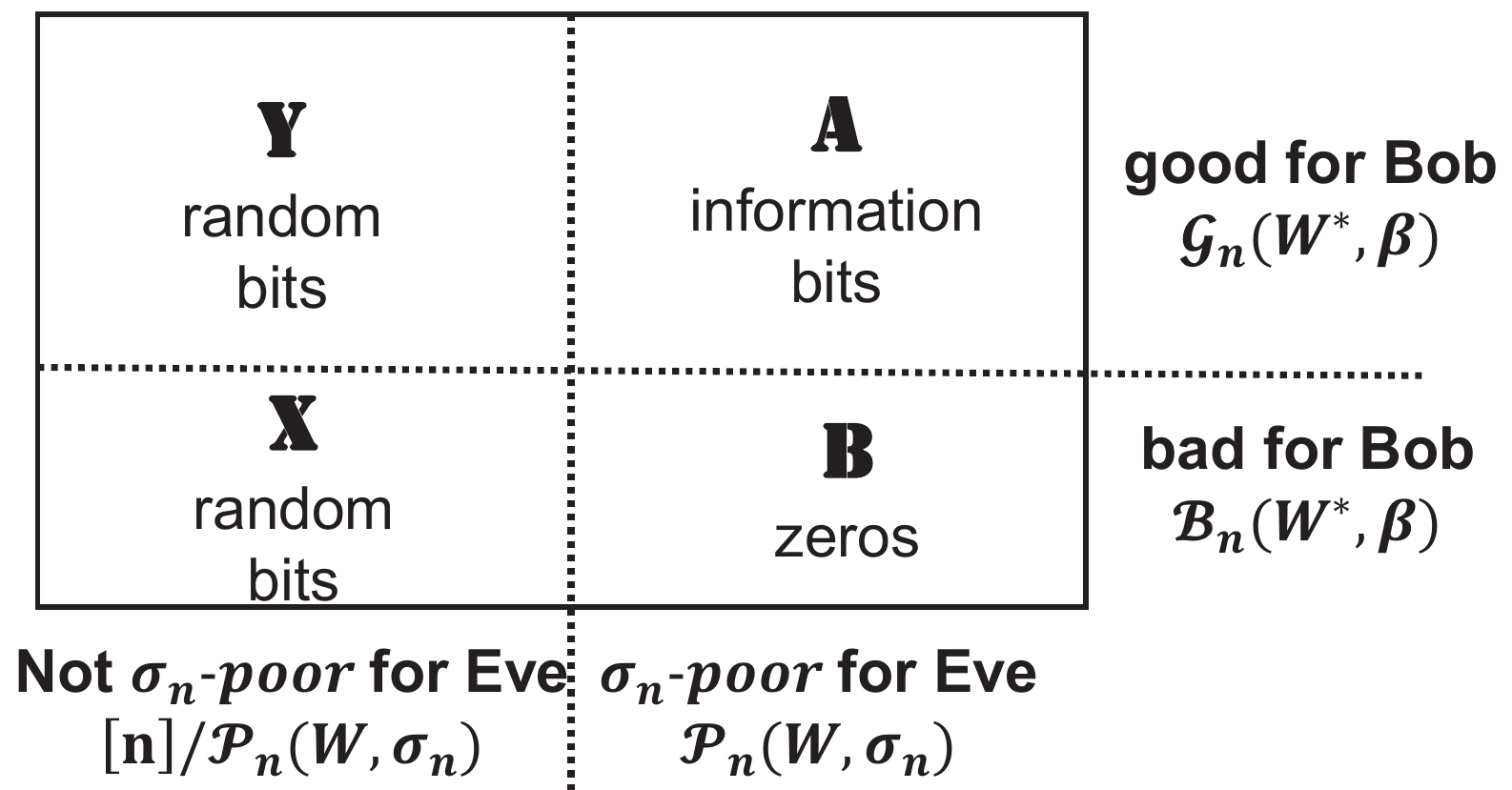}
\caption{Mahdavifar and Vardy's strongly secure channel coding scheme.}
\label{strongsecure-1}
\end{figure}

Suppose that the main channel is $W^*=BSC(p)$, and the wiretapper's channel is $W=BSC(q)$.
Define index sets $\mathbf{A}$, $\mathbf{B}$, $\mathbf{X}$, and  $\mathbf{Y}$ as follows:
\begin{eqnarray}
  \mathbf{A} &=& \mathcal{P}_n(W,\sigma_n)\cap\mathcal{G}_n(W^*,\beta) \\
  \mathbf{B} &=& \mathcal{P}_n(W,\sigma_n)\cap\mathcal{B}_n(W^*,\beta) \\
 \mathbf{ X} &=& \{[n]\setminus \mathcal{P}_n(W,\sigma_n)\}\cap\mathcal{B}_n(W^*,\beta) \\
  \mathbf{Y} &=& \{[n]\setminus \mathcal{P}_n(W,\sigma_n)\}\cap\mathcal{G}_n(W^*,\beta)
\end{eqnarray}
The strongly secure channel coding scheme proposed Mahdavifar and Vardy is shown in Fig. \ref{strongsecure-1},
in which the channels in  $\mathbf{A}$ are used to transmit information bits; the channels in $\mathbf{B}$ are used to send zeros; and channels in $\mathbf{X}$ and $\mathbf{Y}$ are used to transmit random bits.
Mahdavifar and Vardy show that the their coding scheme is secrecy capacity achievable with strong security in \cite{Mahdavifar2011}.
The proposed encoding and decoding process is described as follows.
\begin{itemize}
  \item {Encoding: Let $\mathbf{u}$ be the information bits in  $\{0,1\}^{|\mathbf{A}|}$.
         Alice selects $\mathbf{e}$ for $\{0,1\}^{|\mathbf{Y}|}$ uniformly at random.
         Taking $v^n(\mathbf{A})=\mathbf{u}$, $v^n(\mathbf{X}\cup\mathbf{Y})=\mathbf{e}$ and $v^n(\mathbf{B})=\mathbf{0}$, the codeword of $v^n$ can be expressed as
\begin{equation}\label{codeword}
  x^n=v^nP_nG^{\otimes r},
\end{equation}
where $v^n(\mathbf{D})=(v_{i_1},v_{i_2}\cdots,v_{i_{|A|}})$ for any index set $\mathbf{D}=\{i_1,v_{i_2},\cdots,i_{|\mathbf{D}|}\}$.}
  \item {Decoding: After receiving $y^n$ from the main channel $W^*$, Bob produces a vector $\hat{v}^n$ by invoking successive cancellation decoding in \cite{Arikan2009} for polar code $\mathcal{C}_n\mathbf{(A}\cup \mathbf{Y})$.}
\end{itemize}

Without loss of generality, we assume that Eve knows the information sets and frozen sets. This can be done by calculating Bhattacharyya parameters if the crossover probability of the main channel is known to Eve.
Clearly, this assumption is reasonable and even strengthens Eve's capability.

In our polar code construction, we take
\begin{equation}\label{sigman}
  \sigma_n=2^{n^{-\gamma}}
\end{equation}
to obtain  $ \sigma_n$-poor bit channels (i.e., Equation (\ref{eq: poorset})) for wiretapper's channel.
From \cite[Theorem 17]{Mahdavifar2011}, this coding scheme is strong security.

As mentioned in \cite{Mahdavifar2011}, it cannot provide the reliability of the main channel when the coding scheme attempts to achieve the secrecy capacity.
However, it is unnecessary to achieve the secrecy capacity when the coding scheme is used in the proposed authentication protocol, as the authentication rate is main determined by the rate of the hash functions.
In fact, we will show that the reliability of the main channel is achievable if the secrecy rate is lower than the secrecy capacity in the following section.

\section{Performance Evaluation}\label{sec:performence}

In this section, we conduct extensive simulations to evaluate the performance of the message authentication protocol over BSWC.
In the simulations, we use LFSR-based $\varepsilon$-AWU$_2$ Hash Functions and strong secure polar codes in the proposed authentication framework.

\subsection{Performance  of Secure Polar Codes}
Since secure polar codes play a key role in the proposed authentication protocol.
It is necessary to evaluate the reliability and security of the channel coding scheme.

Taking $\beta=0.1$, $\gamma=0.1$, and the main channel's crossover probability $p=0.1$, an experiment is designed  to test the decoding error with respect to different code lengthes as follows.  Alice encodes a randomly chosen message with coding scheme in Section \ref{polarcode}, and then, transmits the codeword over wiretap channel (BSC($p$), BSC($q$)); after receiving the output of their respective channels, Bob and Eve invoke successive cancellation decoding algorithm in \cite{Arikan2009}.
Here, we repeat the experiment 100 times for each set of parameters.

\begin{figure}[!t]
\centering
\includegraphics[scale=0.6]{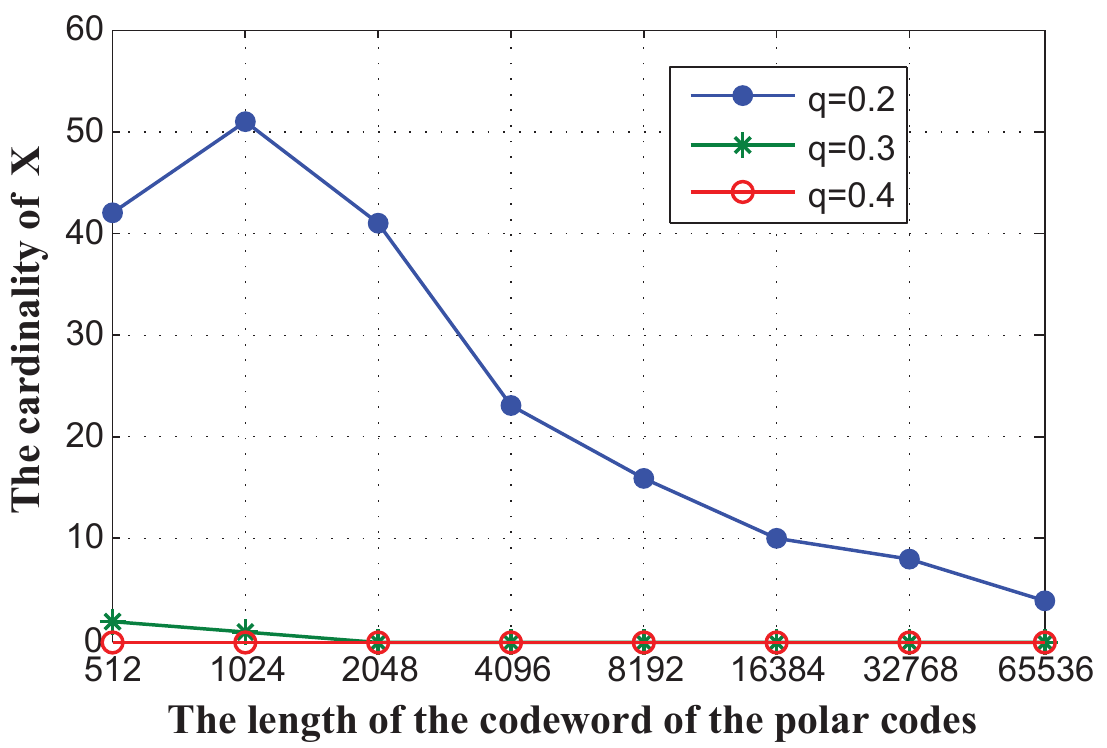}
\caption{The cardinality of index set $X$.}
\label{sizeofX}
\end{figure}

From the simulations above, we find that all the decoding error rates at Bob is zero, i.e., the polar codes constructed by in Section \ref{polarcode} with the parameters setting above can be correctly decoded by the main user.
Note that, the reliability of the coding scheme mainly depends on the the size of index set $\textbf{X}$.
Fig. \ref{sizeofX} shows the cardinality of $\textbf{X}$ with respect to the code length $n$.
It can be seen that the number of $\mathbf{X}$ for each $n$ under different wiretap channel scenarios are small.
As a result, the reliability of the coding scheme is achievable.

\begin{figure}[!t]
\centering
\includegraphics[scale=0.62]{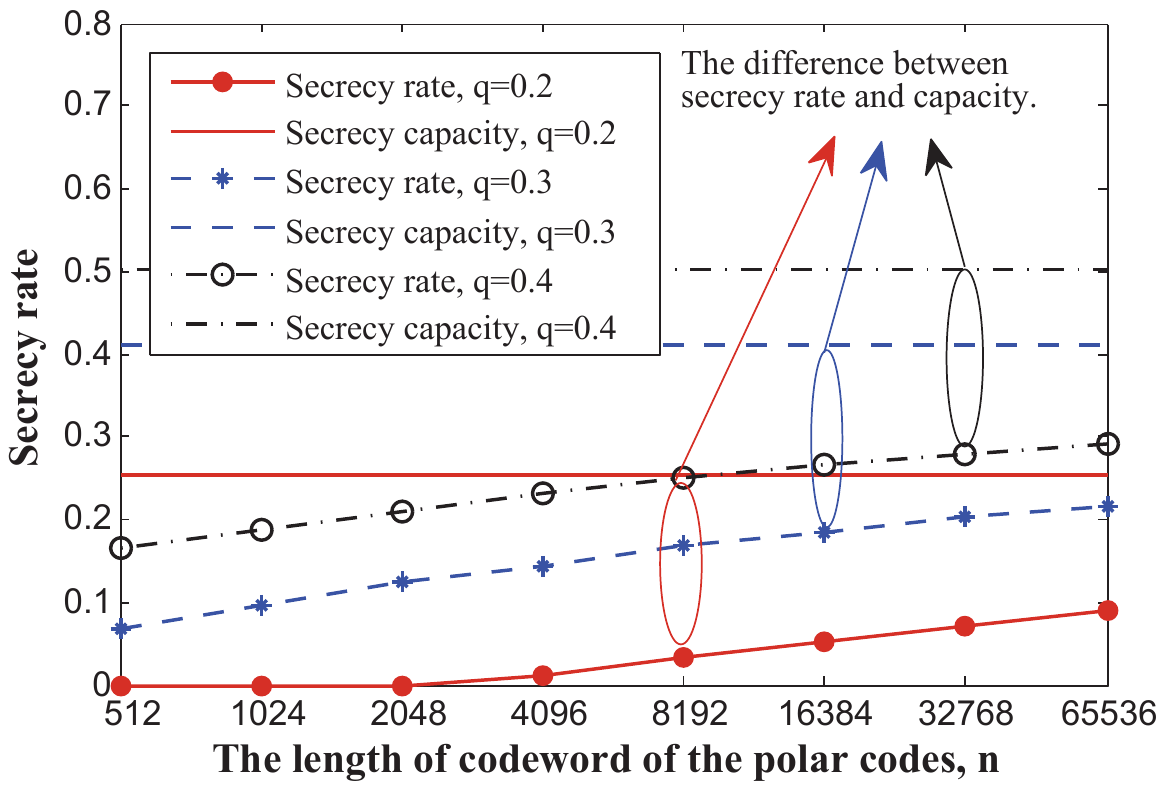}
\caption{Secrecy rate versus secrecy capacity.}
\label{capacityvsrate}
\end{figure}

Fig. \ref{capacityvsrate} compares the secrecy capability of the corresponding wiretap channel and the secrecy rate of the polar codes under different code lengthes.
It can be seen that 1) a larger code length improves the secrecy rate; and 2) there is a substantial gap between secrecy rate and secrecy capacity.
The results imply that, 1) a larger code length can lead to a higher authentication rate when polar codes is used in the proposed authentication protocol; and 2) there is substantial gap between the code rate and channel capacity, which leads to the reliability of the coding scheme, as a lower code rate results in a higher reliability.

\begin{figure}[!t]
\centering
\includegraphics[scale=0.42]{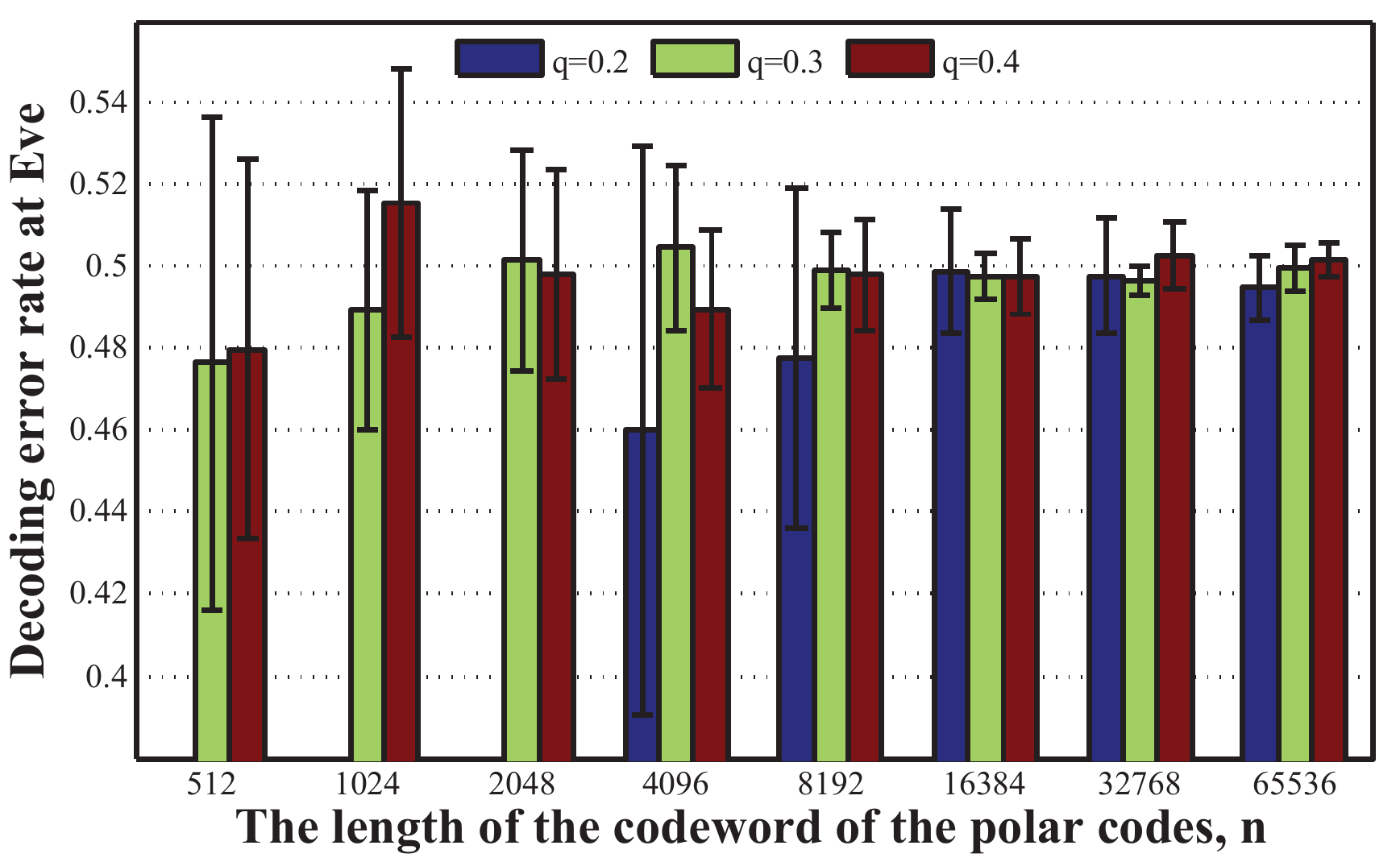}
\caption{Decoding error rate at Eve.}
\label{ErrorRateatEve}
\end{figure}

Fig. \ref{ErrorRateatEve} shows the decoding error rate at Eve versus the code length under different set of parameters and wiretap channel scenarios.
It can be seen that the error rate is closer to 0.5 when the code length $n$ decreases.
It is worth noting that, a smaller  absolute value of the difference between the error rate and 0.5 means
a higher entropy of the secure information at Eve, which further indicates  a higher security of the secure information.
As shown in this figure, the bar of decoding error rate  does not exist for $q=0.2$ and $n=512$ (or $n=1024$).
The reason is that, $\mathbf{A}=\emptyset$ in these cases, which means Alice cannot transmit any secure information to Bob.

\begin{figure}[!t]
\centering
\includegraphics[scale=0.62]{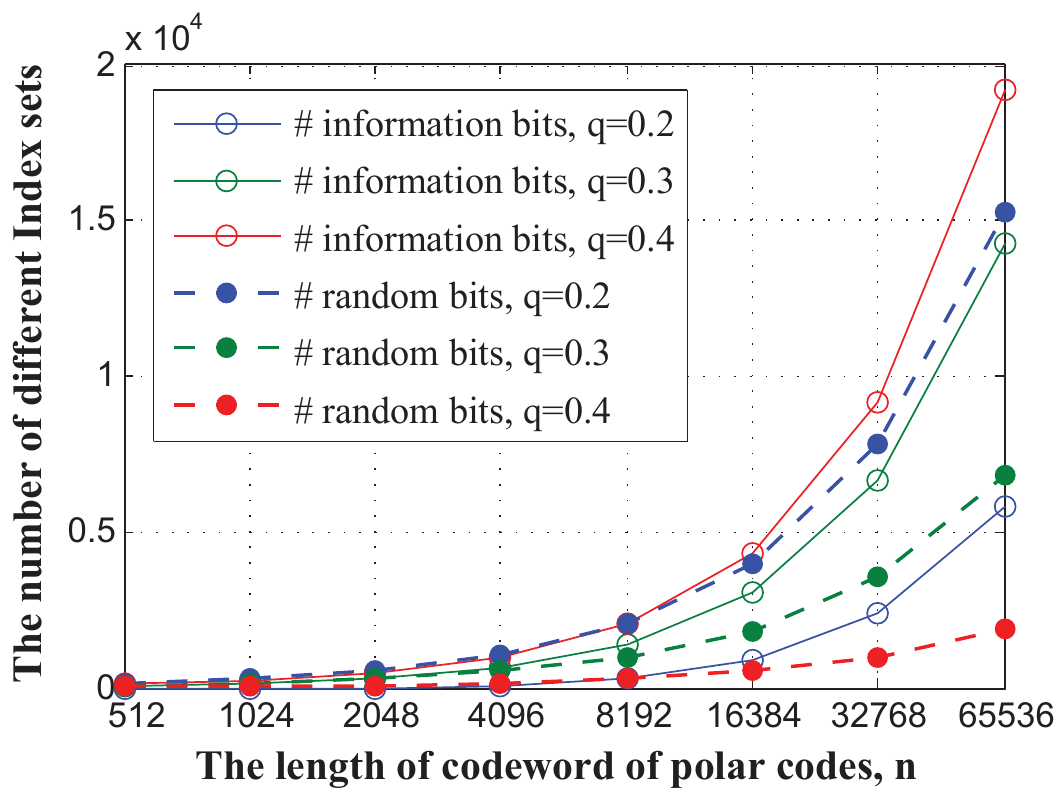}
\caption{The number of different index sets with varying code length.}
\label{Indexsetnumber}
\end{figure}

Fig. \ref{Indexsetnumber} shows the number of information bits and random bits versus the code length under different wiretap channel scenarios.
We can find that both the number of information bits and random bits increase with the increasing of code length.
Therefore, for any length $|S|$ of the message tag, there exists a code length $n_0$, such that $|S|\leq{|\textbf{A}(n)|}$ for any $n>n_0$, where $\textbf{A}(n)$ is the index set of information bits in terms of code length $n$.

\subsection{Exploration of Protocol Parameters}

\begin{figure}[!t]
\centering
\includegraphics[scale=0.58]{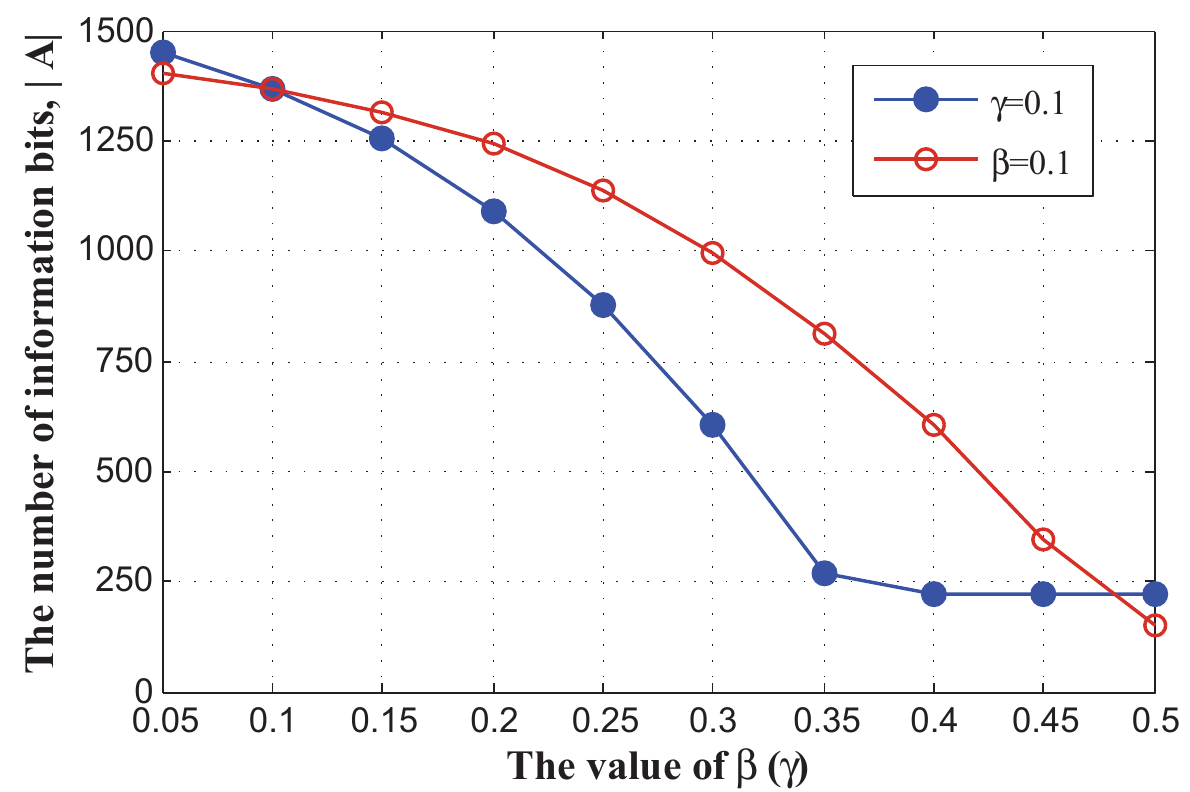}
\caption{The number of information bits against $\beta$ and $\gamma$.}
\label{Numinformationbits}
\end{figure}

\begin{figure}[!t]
\centering
\includegraphics[scale=0.6]{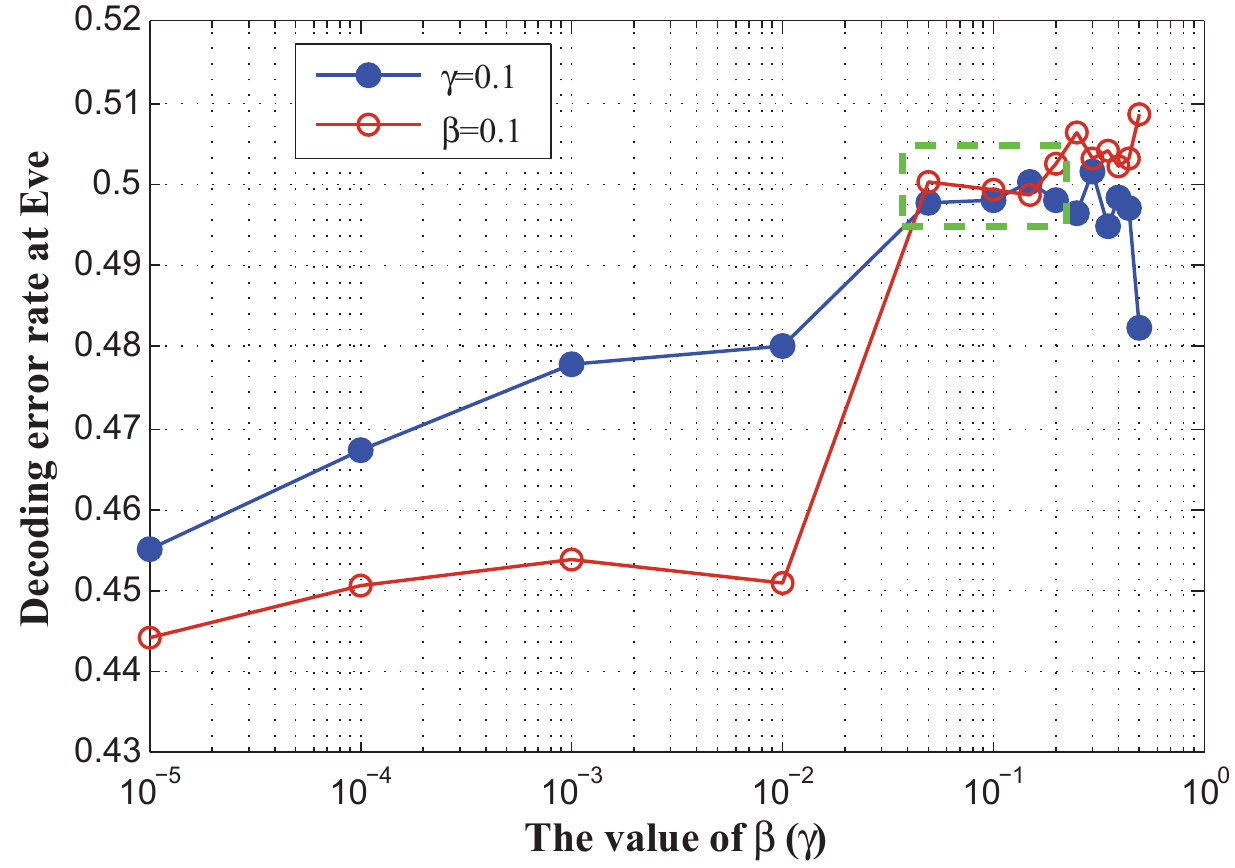}
\caption{The decoding error rate against $\beta$ and $\gamma$.}
\label{ErrorateveBG}
\end{figure}

We now consider the effect of the channel partition parameters $\beta$ and $\gamma$ on the proposed protocol.
Fig.\ref{Numinformationbits} and Fig. \ref{ErrorateveBG} show the change of the number of information bits and the decoding error rate at Eve, respectively, versus varying parameters $\beta$ and $\gamma$, where $n=8192$, $p=0.1$, and $q=0.3$.

As shown in Fig. \ref{Numinformationbits}, the blue line corresponds to the number of information bits by changing the value of $\beta$ and fixing $\gamma=0.1$, while the red line describes the number of information bits by fixing $\beta=0.1$ and varying the value of $\gamma$.
The result shows that the number of the information bit decreases with increasing the $\beta$ and $\gamma$.
The reason is that the cardinality of $\mathcal{P}_n(W,\sigma_n)$ and $\mathcal{G}_n(W^*,\beta)$ decrease with increasing $\beta$ and $\gamma$, respectively.
From this result, a smaller value of $\beta$ and $\gamma$ means a longer length of authentication tag, and further implies a higher authentication rate.

As shown in  Fig. \ref{ErrorateveBG}, the blue line corresponds to the the decoding error rate at Eve by changing the value of $\beta$  and fixing $\gamma=0.1$, while the red line describes then  decoding error rate at Eve by fixing $\beta=0.1$ and varying the value of $\gamma$.
It can be seen that, the error rate is less than $0.48$ for $\beta,\gamma\in(0,0.05)$; and the error rate is in $[0.495,0.505]$ for $\beta,\gamma\in[0.05,0.2]$.
The reason is that, if $\beta$ and $\gamma$ are too small, the size of the set that is both good for Bob and poor for Eve will be large. As a result, the probability of the event that, the indexes which are ``not so bad'' for Eve are in $\mathbf{A}$, will increase.

Based on the discussion above, for both efficiency and security of the proposed protocol, we can choose $\beta$ and $\gamma$ in  the interval $[0.05,0.2]$.

\subsection{Performance of the Proposed Protocol}
We conduct simulations in a large variety of paraments under different crossover probabilities in both main channel and wiretap channel. We use the metrics for performance evaluation as follows:
time cost,  authentication error rate, and authentication rate.

Let the message length  be $2^{\eta}$ bits, and the tag length be $\eta'$ bits,
The family of hash functions is
\begin{equation}\label{examplehashing}
\left\{\Psi_{k}:\{0,1\}^{2^{\eta}}\rightarrow \{0,1\}^{\eta'}\right\}_k,
\end{equation}
where $k=(p(x),e,b)$, and $e$ is the initial state of the LFSR with primitive generator polynomial $p(x)$.
So, the key length is $3\eta'$ bits and $\varepsilon\leq 2^{\eta-\eta'+1}$.

In the authentication protocol, we require that the index set number is larger than the tag length, i.e., $|\mathbf{A}|\geq \eta'$.
If  $|\mathbf{A}|>\eta'$, Alice transmits message tag $s$ with the first $\eta'$ channels in $\mathbf{A}$, and  sends random bits with the remaining channels in $\mathbf{A}$.

\begin{figure}[!t]
\centering
\includegraphics[scale=0.58]{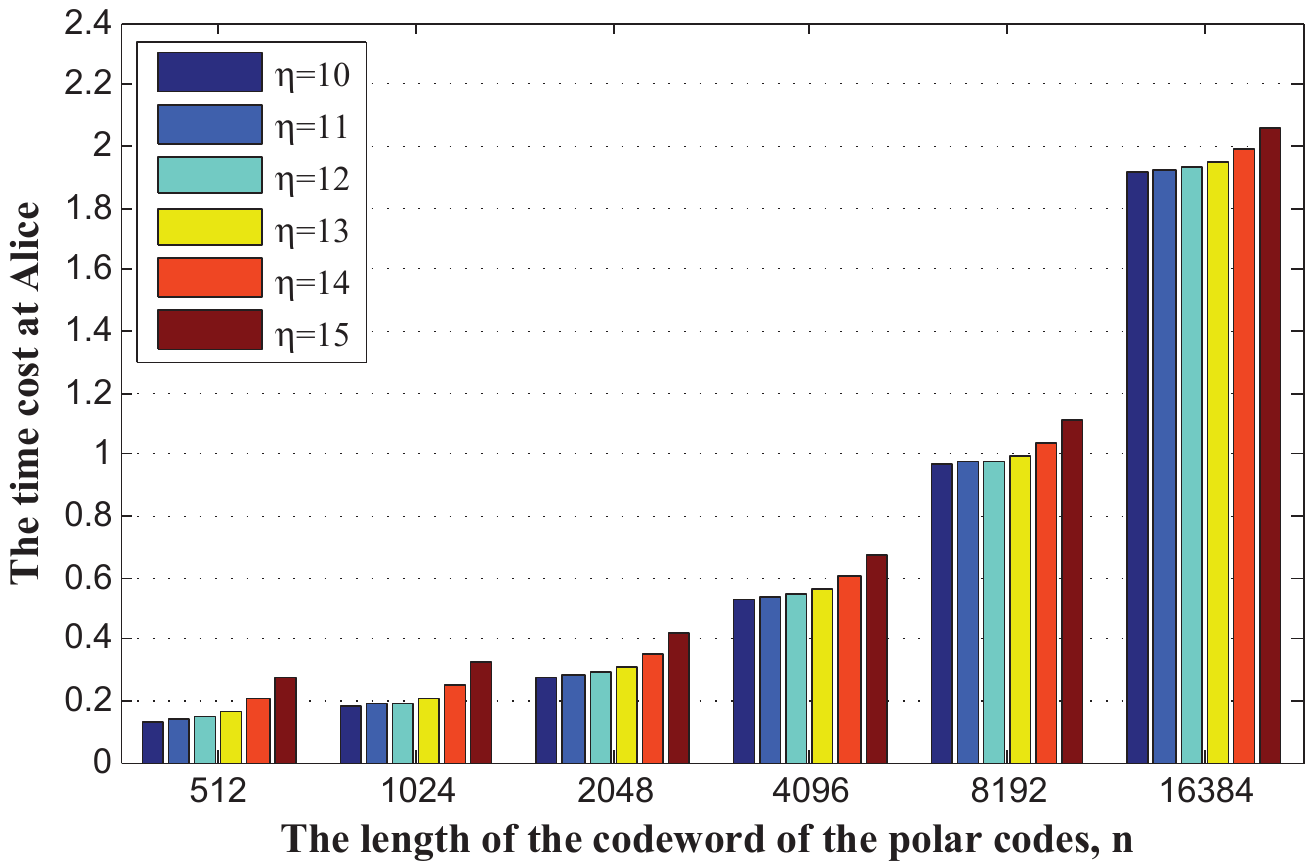}
\caption{Time cost at Alice.}
\label{timecostAlice}
\end{figure}

\begin{figure}[!t]
\centering
\includegraphics[scale=0.58]{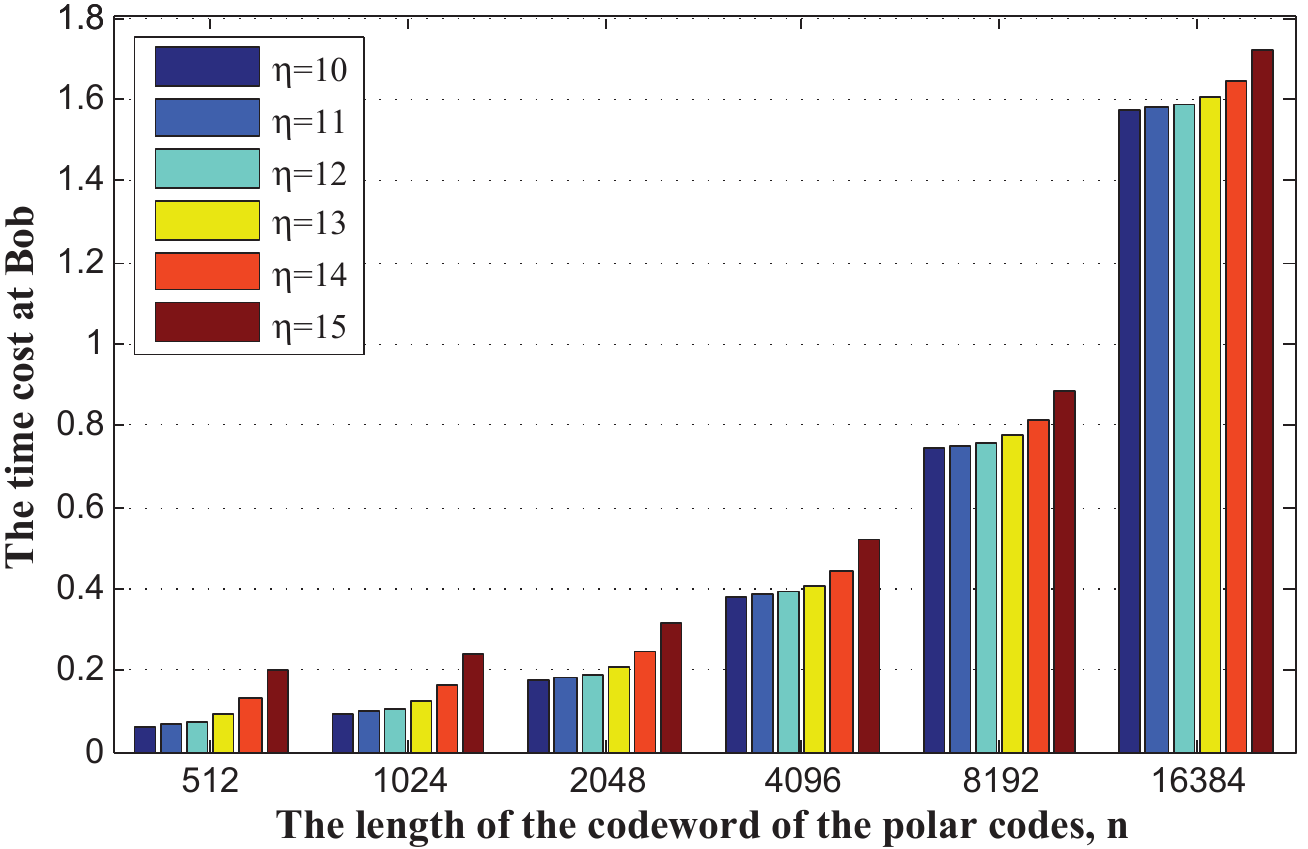}
\caption{Time cost at Bob.}
\label{timecostBob}
\end{figure}

We first consider the time cost of the proposed protocol.
The time cost at Alice includes the tag generation time and the encoding time, while the time cost at Bob involves the tag generation time and the decoding time.
Fig. \ref{timecostAlice} and Fig. \ref{timecostBob} show the time cost at Alice and Bob, respectively,  with respect to the code length with different message lengths. Here $\beta=0.1$, $\gamma=0.1$, $q=0.1$, $p=0.3$,  the message length is  $2^\eta$, and
the primitive generator polynomial used to generate the LFSR is:
\begin{equation}\label{primeploy101}
 p(x)=x^{101}+x^{84}+x^{66}+x^{49}+x^{32}+x^{16}+1.
\end{equation}
The results show that 1) a larger message length and code length improve the time cost;
and 2) the impact of  the code length on time cost is more sensitive than that of the message length on time cost.
The latter result further reveals that the encoding/decoding time is far larger than the hashing time.
Fortunately, the time cost of the proposed protocol will dramatically decrease when this protocol is implemented in hardware, as the encoding/decoding process of polar code and LFSR can be implemented in hardware with a low time cost.


\begin{figure}[!t]
\centering
\includegraphics[scale=0.68]{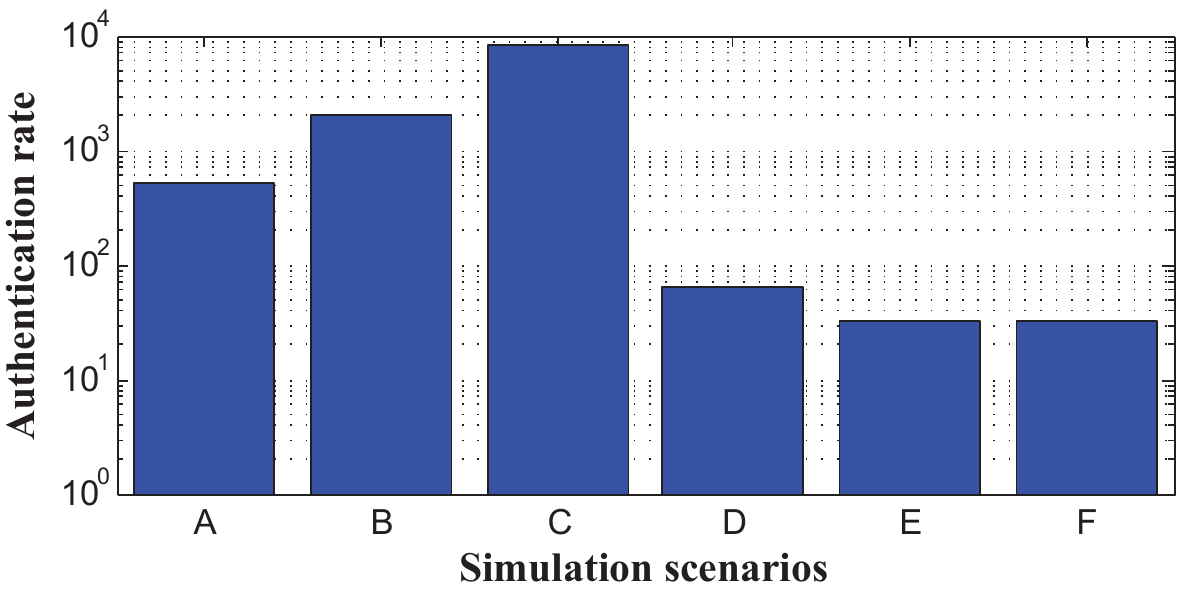}
\caption{Authentication rate under different scenarios.}
\label{Authrate}
\end{figure}

\begin{table}[h]
\caption{Simulation scenarios}
\label{table:revisedprotocol}      
\centering
\small{
\begin{tabular}{cccccc}
\hline\noalign{\smallskip}
  $Index$ &p&q &$L(M)$  & $L(K)$& $L(S)$\\
\noalign{\smallskip}\hline\noalign{\smallskip}
A & 0.1& 0.2 & $2^{25}$& 303 bits & 101 bits\\
B & 0.1& 0.3 &  $2^{25}$& 303 bits & 101 bits\\
C & 0.1& 0.4 &  $2^{25}$& 303 bits & 101 bits\\
D & 0.2& 0.3 &  $2^{20}$& 192 bits & 64 bits\\
E & 0.2& 0.4 &  $2^{20}$& 192 bits & 64 bits\\
F & 0.3& 0.4 &  $2^{20}$& 192 bits & 64 bits\\
 \hline
\end{tabular}
}
\end{table}
Then, we evaluate the authentication error rate and authentication rate of the proposed protocol under different scenarios.
In this simulation, we set $\beta=0.1$, $\gamma=0.1$ and $n=8192$, and take the primitive generator polynomial to generate the LFSR as Equation (\ref{primeploy101}) and
\begin{equation}\label{primeploy}
 p(x)=x^{64}+x^9+x^8+x^7+x^6+x^3+1
\end{equation}
for tag length $L(S)=101$ and $L(S)=64$, respectively.
The remaining parameters for each scenario are listed in Table \ref{table:revisedprotocol}, where $L(M)$ is the message length, and $L(K)$ is the key length.
For each scenario, we repeat the experiment 100 times.
We find that all the messages authenticated by Alice is accepted by Bob in the simulations,
i.e., the authentication error rate is zero in the simulations.
Fig. \ref{Authrate} shows the authentication rate for each scenario.
It can be seen that the proposed protocol can achieve a high authentication rate.

\section{Conclusion}
\label{sec:conclusion}
In this paper, we have proposed a multi-message authentication framework, over Wiretap Channel, to achieve information-theoretic security using the same key. The proposed framework bridges the two research areas in PHY-layer security: secure transmission and message authentication. Moreover, we have developed a theorem revealing the requirements/conditions for information-theoretic security, which provide guidance and insights for authentication protocol design. We have further devised a multi-message authentication protocol to meet the security requirements, with high efficiency. The theoretical analysis demonstrated that the proposed protocol is information-theoretic secure for a polynomial number of messages and attacks. Furthermore, the authentication rate of the proposed protocol approaches infinity when $n$ goes to infinity.
Finally, an efficient and feasible authentication protocol over binary symmetric wiretap channel has been proposed,  and experiments results showed  that this protocol has a good performance.
For the future work, we will extend this study to develop a computationally efficient protocol for Gaussian wiretap channels.

\section*{Acknowledgments}
 \noindent 
 This work is jointly supported by Natural Sciences and Engineering
Research Council (NSERC) of Canada, NSFC (No. 61502085, No. 60973161, No. 61133016), and China Postdoctoral Science Foundation funded project (No. 2015M570775).

\bibliographystyle{acm}

{

}

\end{document}